\RequirePackage{fix-cm}
\documentclass[smallextended]{svjour3}       
\smartqed  
\usepackage{graphicx}
\journalname{Multimedia Tools and Applications}

\usepackage{subfigure}
\usepackage{multirow}
\usepackage{float}
\usepackage{booktabs}
\usepackage{hyperref}
\usepackage[section]{placeins}
\usepackage{amsmath}

\begin{document}

\title{Context-aware adaptation of mobile video decoding resolution\thanks{The research presented in this paper was funded by Slovenian Research Agency (grants no. P2-0098 and N2-0136 ``Bringing Resource Efficiency to Smartphones with Approximate Computing").}
}


\author{Octavian Machidon \and Jani Asprov \and Tine Fajfar \and Veljko Pejović}


\institute{O. Machidon, J. Asprov, T. Fajfar, V. Pejović  \at
              Faculty of Computer and Information Science \\
              University of Ljubljana, Slovenia \\
              \email{octavian.machidon@fri.uni-lj.si, ja0905@student.uni-lj.si, tf5442@student.uni-lj.si, veljko.pejovic@fri.uni-lj.si}
}

\date{Received: date / Accepted: date}

\maketitle

\begin{abstract}
While the evolution of mobile computing is experiencing a considerable growth, it is at the same time seriously threatened by the limitations of the battery technology, which does not keep pace with the evergrowing increase in energy requirements of mobile applications. A novel approach for reducing the energy appetite of mobile apps comes from the approximate computing field, which proposes techniques that in a controlled manner sacrifice computation accuracy for higher energy savings. Building on this philosophy we propose a context-aware mobile video quality adaptation that reduces the energy needed for video playback, while ensuring that a user's quality expectations with respect to the mobile video are met. We confirm that the decoding resolution can play a significant role in reducing the overall power consumption of a mobile device and conduct two user studies to investigate how the context in which a video is played, it's content, and the user's personality, modulate a user's quality expectations. We discover that a user's physical activity, the spatial/temporal properties of the video, and the user's personality traits interact and jointly influence the minimal acceptable playback resolution, paving the way for context-adaptable approximate mobile computing.
\keywords{Mobile computing \and Approximate computing \and Video decoding \and Context inference \and Spatial information \and Temporal information}

\end{abstract}

\section{Introduction}
\label{sec:introduction}

Mobile computing has been experiencing an overwhelming expansion in the last few decades, with the smartphone -- which was invented only slightly more than a dacade ago -- being owned today by more than three billion people (3.6 billion users in 2020, 4.3 billion users forcasted for 2023 \cite{smartphoneStatista}). In today's world, mobile computing has become ubiquitous, and the mobile applications and wireless technologies transformed the way we communicate, do business, navigate in space, or find social contacts. 

One of the staggering changes fostered by the proliferation of mobile computing and the technological advances in smartphone technology is in the way information is consumed on mobile devices, with the focus moving from the traditional voice and text media to video content. Surveys show that already 90\% of the owners watch videos on their mobile devices and that more than 70\% of all YouTube content is consumed via mobile devices \cite{youtube2020}. The amount of content seen through mobile video is more than doubling every two years~\cite{forecast2019cisco}.  In 2019, mobile video traffic accounted for half of the total mobile data traffic and the forecast indicates that almost 80\% of the worldwide mobile data traffic will be video traffic by 2022 \cite{forecast2019cisco}. This growth in mobile video streaming has been further exacerbated recently by the COVID-19 pandemic, with the fields as diverse as the education, remote work, and healthcare, rapidly jumping on the mobile video bandwagon~\cite{ForbesCovid}. 

Nevertheless, the proliferation of mobile computing in general, and even more specifically of mobile video streaming, is hindered by the physical constraints and limitations of the underlying hardware. One key issue in this regard is related to one of the most critical resources of a mobile device -- its battery. Mobile video streaming applications are among the most power-hungry smartphone apps~\cite{Hoque2013} and the intensive growth in the amount of mobile video streaming data continues to put significant pressure on the power consumption of smart mobile devices~\cite{Zou2017}. At the same time, the battery technology is experiencing a disproportionally slower growth -- practically a stagnation –- compared to the other mobile resources including the CPU speed and computing power, storage space, and wireless transmission speed~\cite{1401839}. The lack of a revolutionary solution for modest battery capacity calls for further efforts towards the efficient use of limited resources available on mobile devices.

Inspired by approximate mobile computing (Section \ref{sec:background:amc}), we investigate the feasibility of implementing context-, content-, and user-dependent video quality adaptation with the goal of improving the energy efficiency of mobile video playback. Our work is driven by the following hypotheses:
\begin{enumerate}
    \item Video playback resolution represents a suitable ``knob'' for trading off video playback quality and the corresponding energy usage;
    \item A viewer's requirements with respect to the video playback quality vary with the physical context (i.e. the activity state) of the viewer;
    \item A viewer's requirements with respect to the video playback quality depend on the content-related properties of the video;
    \item Subjective factors pertaining to the viewer may influence the required video quality.
\end{enumerate}

We start by performing fine-grain energy measurements to confirm the first hypothesis (Section \ref{sec:background:tradeoff}). We then conduct two studies, described in Section~\ref{sec:methods}, to examine the remaining three assumptions. The first study is targeted on investigating the influence of contextual situations (such as whether a user is still, running, walking, or riding in a car) on the video quality requirements. The study confirms that these factors significantly impact the minimum playback resolution the user is satisfied with. In addition to this, findings further examined in Section~\ref{sec:results} uncover other aspects that can also play a role in the user's tolerance with lower video quality, such as the video's content (described by its spatial and temporal complexity) and user-related factors, confirming the last two assumptions. Building upon these initial findings, we design the second study more rigorously targeted to investigate the impact of the video's spatial and temporal characteristics on the required playback quality. In addition, we also examine other human factors that could influence user quality expectations, such as the user's personality traits. Thus, in second study we also collect information on the personality of the participants, more in-depth information about the properties of the video content, and employ a more rigorous statistical analysis based on mixed linear models. Our investigation clearly pinpoints the physical activity, but also the interplay between the physical activity and the video content, as well as the impact of personality and gender-related factors on the opportunities for reducing the mobile video energy requirements through controlled approximation. 

Merging mobile systems design, mobile sensing, and human-computer interaction, our work opens space for dynamic minimization of the gap between the users needs and the computational effort delivered by mobile computers. Furthermore, the contextual information, including a viewer's mobility state, properties of the video content, and even personality traits, may be acquired with very little cost/overhead in today's ubiquitous mobile devices and apps, thus our work remains readily implementable in practice, providing a new dimension to the existing, mostly statically applied, approaches to resource-efficient multimedia described in Section~\ref{sec:related_work}. Finally, in Section~\ref{sec:conclusions} we discuss future research avenues in the area of mobile video adaptation, but also in the area of approximate mobile computing in general.

This paper represents original work by its authors, yet the initial findings on the relationship between the minimum tolerated video playback resolution and a user's physical activity state were presented in our Mobiquitous 2020 conference paper~\cite{machidon2020watching}. While building upon the general idea of~\cite{machidon2020watching}, in the current manuscript we greatly expand this research by thoroughly investigating how spatial and temporal properties of the video modulate the relationship between the desired resolution and a user's physical activity. Furthermore, we for the first time examine the role of a user's personality aspects on the mobile video resolution requirements. The additional investigations are conducted through a separate user study with 22 users who had not participated in the original study. Finally, we fully revise the statistical methodology that now includes sophisticated hierarchical modelling of the target relationships. Our work was performed with reproductibility in mind and the collected experimental data from both studies is publicly available to the research community at \url{https://gitlab.fri.uni-lj.si/lrk/approximate_video_study/}.

\section{Background \& Preliminaries}
\label{sec:background}

\subsection{Towards Approximate Mobile Computing}
\label{sec:background:amc}

Approximate computing (AC) is a resource-efficient computing paradigm grounded in the observation that the result of a computation often need not be perfectly accurate to satisfy the end-user's needs~\cite{10.1145/2821510}. Opportunities for AC frequently arise when the computation inputs are noisy (e.g. sensor data), or when the output is further manipulated and interpreted by the user (e.g. augmented reality rendering). In such situations, approximate computation can deliver a fully satisfactory result while reducing the energy use. AC techniques have already proven their efficiency in various desktop scenarios, with approaches ranging from speeding up code execution through compiler-level optimizations that omit certain lines of code \cite{6062070} to performing neural-network based approximations instead of complex function calculations~\cite{6493641}, demonstrating significant energy savings while maintaining acceptable result accuracy. 

Building upon the idea of AC, approximate mobile computing (AMC) introduces approximation on mobile devices~\cite{Pejovic2019AMC}. The core difference from the conventional AC being the context of use, which in mobile computing tends to vary over time. A user's physical activity, location and collocation with other users, the outside brightness, and numerous other factors may vary throughout the day and impact the user's requirements with respect to mobile computation. Significant challenges lay ahead before the full potential of AMC can be exploited: 1) practical means of enabling approximation in mobile apps need to be provided; 2) the benefits of approximate execution need to be quantified; 3) opportunities for approximation need to be identified and profiled, and 4) lightweight context recognition relevant for AMC needs to be implemented.

This paper describes our efforts towards enabling AMC in the field of mobile video playback. This field represents not only one of the most prominent aspects of mobile computing, but is also among the most energy hungry ones~\cite{zhang2017}. We hypothesize that the context of the mobile video playback impacts the user's perception and quality requirements. By ``context'' one can understand a potentially unlimited number of dimensions, however, backed by the prior work~\cite{trestian,Song2011,SEETO20121484} in our experiments we focus on the three most relevant and intuitive dimensions -- a user's physical activity, the characteristics of the mobile video, and the user's personality traits. Consequently, we formulate the following research questions (RQ) that our study aims to answer:
\begin{itemize}
    \item \emph{RQ\textsubscript{1}: Does the physical activity the user is engaged in when watching a video on a mobile device influence the user's quality expectations/requirements?}
    \item \emph{RQ\textsubscript{2}: Does the video content (its spatial and temporal characteristics) impact the user's satisfaction with a given video playback quality and does the physical mobility state of the user modulate the relationship between the video content properties and the desired playback quality?}
    \item \emph{RQ\textsubscript{3}: Do the user's personality traits impact the quality requirements of a mobile video playback?}
\end{itemize}
\noindent In addition, we are interested in the potential of enabling energy savings by adjusting video playback according to the current context. Thus, we also aim to answer:
\begin{itemize}
    \item \emph{RQ\textsubscript{4}: How much energy can be saved by employing a predictive context-, content- and personality-aware mobile video resolution model that adjusts video playback quality to the minimal level that still satisfies the user's quality expectations?}
\end{itemize}

To realize AMC the first step is to provide straightforward and efficient means of adjusting approximation. In addition, the reduction in computations (e.g. decreased resolution) should lead to a gradual decrease in the end-result accuracy (e.g. user quality perception), without the loss of correctness (i.e. the result is usable at all times, and the approximation ``knob'' always gives a correct result). Moreover, the reduction in computation should translate to reduced resource usage (and thus energy savings). In our work we settle on \textit{video decoding resolution} adjustment. Virtually all video distribution frameworks (e.g. Youtube, Vimeo), as well as mobile video players, support playback resolution adaptation. Furthermore, setting video resolution always leads to correct execution and the loss of quality is gradual as we dial down the resolution. In the following section we also confirm that the loss of quality corresponds to lower resource usage making video decoding resolution a suitable technique for approximate computing adaptation. 

\subsection{Energy vs. Quality Trade-off in Mobile Video Decoding}
\label{sec:background:tradeoff}

The approximate computing philosophy has at its core the monotonically increasing relationship between the computation accuracy and the resource consumption. In this section we chart the relationship between the video decoding quality and the mobile consumption. When performing the energy measurements, we use a popular video decoding software VLC Player \cite{vlc} running on a Samsung Galaxy S3 (I9300) Android smartphone. Despite being released nine years ago, the phone supports both hardware and software video decoding and, importantly, has a detachable battery that allows us to connect the phone to a high-frequency power meter. The VLC Player was chosen for the energy measurements due to its flexibility in allowing rapid enabling/disabling of hardware accelerated decoding.


The experimental setup for measuring energy consumption relies on measurements from the Monsoon High Voltage Power Monitor (HVPM)~\cite{monsoon}, a high sampling frequency platform commonly used for power measurements in mobile computing~\cite{Schuler2019}. This platform generates energy readings at a sampling frequency of $5 kHz$. Each sample contains a timestamp in $ms$, voltage in $mV$ and electrical current in $mA$. The HVPM is directly attached to the battery interface of the mobile device, which is powered solely by the HVPM. 

During the energy measurements, the HPVM output voltage was set to $4.2 V$ corresponding to the voltage of an almost full battery. The same 1-minute video was downloaded from YouTube on the device in the following resolutions: 144p, 240p, 360p, 480p, 720p and 1080p, in both WebM and MPEG-4 formats. The baseline for comparison was a reference energy measurement performed with just the phone screen turned on, without other apps/services running. For each resolution, the video was played 10 times using VLC Player and the energy readings were averaged over the 10 runs. During the measurements, the screen brightness was set to the minimum, all non-essential services running on the smartphone that could interfere with the energy measurements were shut down, and the smartphone's Airplane mode was turned on to avoid the effect of on-device communication modules (e.g. GSM, Wi-Fi, Bluetooth, etc.).


\begin{figure}
\centering
  \includegraphics[scale=0.6]{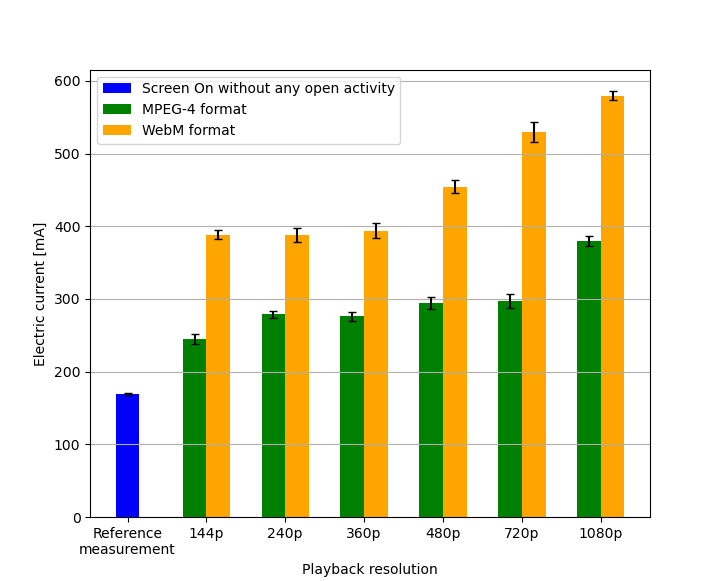}
  \caption{Smartphone average current consumption during video playback at different resolutions together with the standard deviation of the measurements. A monotonically increasing relationship between the video decoding resolution and the current consumption is evident for both software (WebM) as well as hardware (MPEG 4) decoding.}
  \label{fig:energy}
\end{figure}

The results of the energy measurements for video playback on the mobile device at different resolutions are shown in Figure~\ref{fig:energy} (for the reference we also show the measurements with the screen turned on, but no playback running). We observe a significant difference in power consumption for playing videos using MPEG-4 vs. WebM decoding. This is expected since MPEG-4 decoding is hardware-accelerated in modern smartphones, while WebM decoding is performed in software. With both formats we see a generally increasing trend -- the higher the decoding quality (resolution), the higher the consumption is. 
Interestingly, in the WebM case the lower resolutions (144p, 240p and 360p) have similar average current consumption, while the consumption increases considerably as we move to higher resolutions (480p, 720p and finally 1080p). Since there are no significant differences between the lower three resolutions, from the energy efficiency point of view, lowering the resolution under 360p would have no positive impact on energy savings, moreover it would only potentially decrease a user's satisfaction. 

\section{Methodology}
\label{sec:methods}

Video decoding resolution represents a suitable knob for controlling approximation, as confirmed by the energy-quality trade-off described in Section~\ref{sec:background:tradeoff}. Yet, it is unclear where on the trade-off line one should operate in order to satisfy the user requirements, while minimizing the energy use. 

Viewer perception of video playback is shaped by a multitude of factors, including quality of image, location and time availability and choice of content~\cite{papagiannidis2009evaluation}. All these dimensions vary according to the platform and context used for visualization (i.e. a mobile device, which might be on the move, or a desktop device indoors). This in turn influences how the sensory, emotional, and cognitive factors influence the viewer’s engagement level, and ultimately the perception and satisfaction with the viewing experience~\cite{SEETO20121484}. For example, the content type determines the availability of sensory experience and emotional response, and the attention span required. Also, the platform and context impact the attention span, since for example mobile context has a much higher level of outside interruption than fixed/desktop usage. In addition, the outside brightness impacts the contrast of the OLED display preventing a viewer from discerning details in the picture. To summarize, the influencing factors collectively form the \textit{context} which, we hypothesize, impacts viewers' requirements with respect to the video playback resolution.



While there are potentially infinite dimensions to the \textit{context}, certain dimensions have already been proven to impact the video perception. For instance, the perception of content rendered on a mobile handheld device's screen can be impacted by \textit{the physical activity }of the viewer, as the ability to focus and interpret the picture may be disturbed~\cite{6777577,6263265}. We therefore first focus on this dimension, which is also characterised by its practical convenience. 
A user's physical activity can be acquired with the minimal use of the mobile's energy. For instance, in Android OS coarse-grained physical activity (e.g. ``running'', ``walking'', ``in vehicle'', ``still'', etc.) can be acquired using Google Play Services' classifier jointly maintained for all apps on the device. Having in mind that activity detection is used across a range of apps, from navigation, over exercise tracking, to health and wellbeing apps, and that an average user has more than thirty apps installed on her phone~\cite{GoogleIpsos2016}, there is a high probability that activity recognition pipeline would anyway be active and routinely queried by other apps. Consequently, querying this classifier for our purpose would likely incur negligible additional energy cost, which makes the physical activity context perfectly suited for our goal of reducing the energy use.

Besides the physical activity, we also hypothesize that \textit{the content of the video} impacts a user's decision to require a higher or a lower resolution decoding. Content information, too, can be acquired with very little cost as no additional device components need to be powered on. 
Therefore, we further calculate a video's spatial and temporal information and inspect their role on a user's desired video playback resolution.

Finally, in addition to the outside contextual factors (user's physical activity state) and the video content, we hypothesise that other internal user factors play a role. As such, we include in our investigation an additional  dimension represented by the viewer's \textit{personality traits}.

\begin{figure}
\centering
  \includegraphics[scale=0.4]{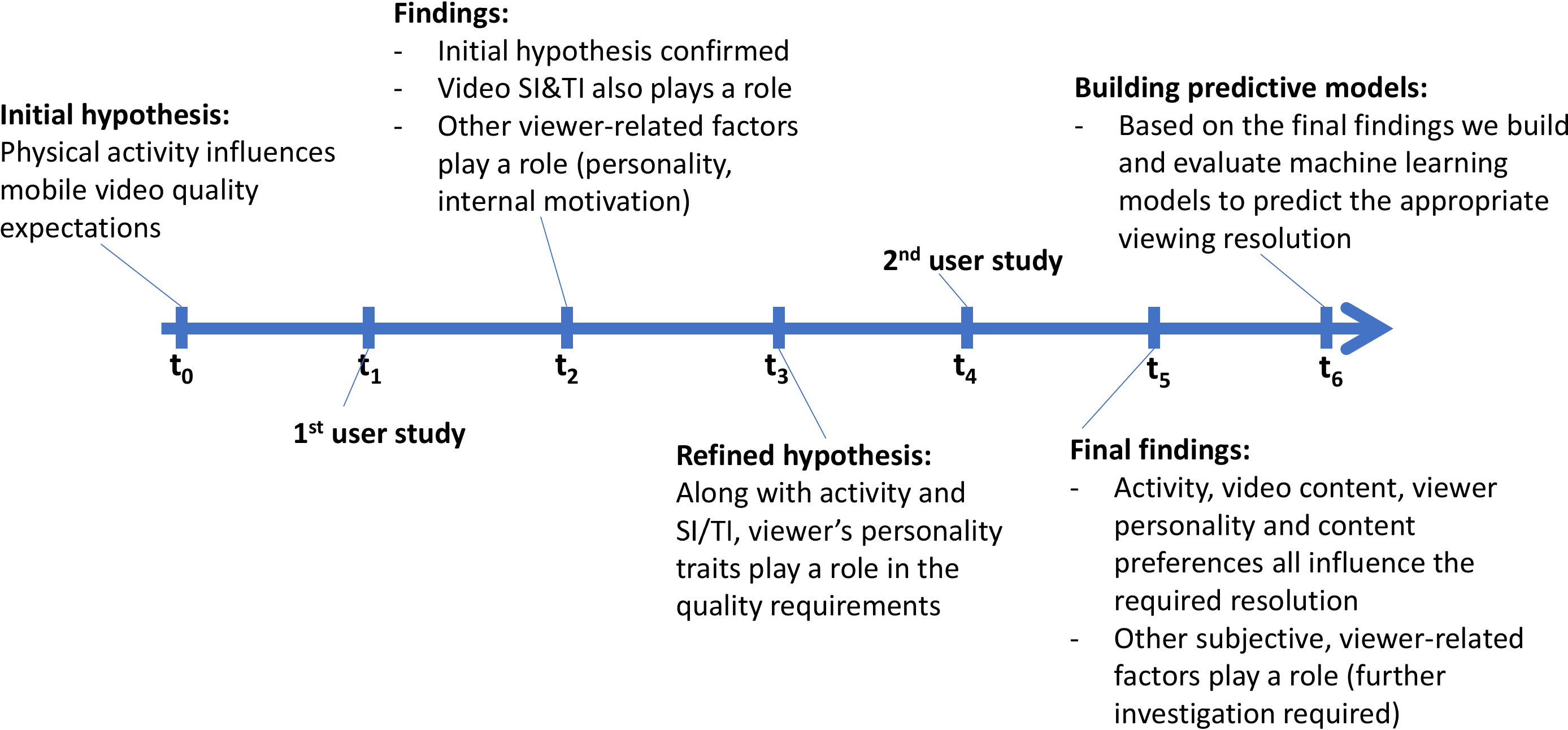}
  \caption{Timeline of the research process, starting from the initial hypothesis, going through the two studies that were conducted, and concluding with the final findings.}
  \label{fig:timeline}
\end{figure}

The outline of the entire research process is illustrated in Figure~\ref{fig:timeline}. We first start our investigation from the hypothesis that physical activity impacts the quality requirements of mobile video rendering. We conduct the first study which confirms this hypothesis, but also indicates that the video content (more specifically, it's spatial and temporal complexity) also plays a significant role in the end resolution required by the viewers. However this study reveals that other viewer-related factors might be important. As such, we conduct the second study, which focuses on the influence of the viewer's personality on the quality expectations. The results of this second study confirm that personality impacts the quality requirements, in addition to the viewer's interest for the content of the video. Finally, the second study also reveals that a significant amount of influence is exerted by other subjective factors, which will require future investigation. Based on these statistical findings, we conclude by building and evaluating machine learning models for predicting the appropriate viewing resolution. High-accuracy models are crucial, should we wish to implement a real-world video adaptation on a mobile device.

\subsection{Mobile Video Management Application}

For video rendering during the user experiments we use NewPipe -- an open source YouTube-streaming frontend for Android \cite{newpipe} -- which allows both online and offline video playing. We choose this app due to its simplicity of use and also flexibility -- being open source it allows us to quickly add new functionalities needed for our experiments. For the scope of the two user studies we conducted, the videos were preloaded to avoid any networking effects that might impact the user perception when watching the videos. We add logging functionalities to the app, thus in each experiment we record the initial resolution, physical activity state, the video played, and each event of a user changing the resolution. For resolution change events we record the new resolution and the timestamp marking the moment the change took place. In this paper we describe controlled experiments, where the users were instructed to perform a certain activity at a certain time, so we could acqure a stratified dataset. Thus, we do not use on-device classifier for recognising activities, but log them manually. Yet, we have also implemented automatic activity recognition and plan to run an in-the-wild study as a part of our future work on automatic resolution adaptation.

\subsection{Video content analysis metrics}

To assess the influence of video content on user satisfaction in different mobility states, for each video we computed two metrics: the average Spatial Information (SI) and the average Temporal Information (TI) indices~\cite{itu}. SI represents the spatial detail in a video frame (complexity) while TI relates to the amount of temporal changes in a video scene (motion), and the two metrics are used for objective video quality prediction \cite{8905186}. The perceived quality of the video after passing through a given digital compression system is a function of the input scene: the amount of motion and spatial detail in a scene correlated with the compression rate of the video influences how the quality of the video is being perceived (e.g. for the same compression rate, a scene with limited motion and spatial detail will be perceived to have higher quality compared to a scene with a large amount of motion and spatial detail, which will appear to be distorted)~\cite{webster1993objective}. 

SI is based on the Sobel filter. Each video frame (luminance plane) at time $n$ ($F_{n}$) is first filtered with the Sobel filter $[Sobel(F_{n})]$. The standard deviation over the pixels ($std_{space}$) in each Sobel-filtered frame is calculated. This step is repeated for each frame in the video sequence and results in a time series of spatial information of the scene. The maximum value in the time series ($max_{time}$) is chosen to represent the spatial information content of the scene \cite{itu}. This process is described by the following equation:

\begin{equation}
    SI=\underset{time}{max} \left \{ std_{space} \left [ Sobel(F_{n}) \right ] \right \}
\end{equation}

TI measures temporal changes (motion) in a sequence of video frames \cite{itu}. TI is based on motion differences between the pixels in the luminance plane of two consecutive frames $F_{n}(i, j)$ and $F_{n-1}(i, j)$, i.e., discrete time $n$ and $n-1$, at pixel position $(i, j)$:

\begin{equation}
    M_{n}\left ( i,j \right )=F_n(i,j)-F_{n-1}(i,j)
\end{equation}

TI is defined as the maximum value of the standard deviations obtained for the sequence of motion differences in the spatial domain~\cite{itu}:

\begin{equation}
    TI=\underset{time}{max}\left \{ std_{space}[M_n(i,j)] \right \}
\end{equation}

\subsection{User Study 1: Mobility State vs Video Resolution Requirements}

\begin{figure}
\centering
  \includegraphics[scale=0.5]{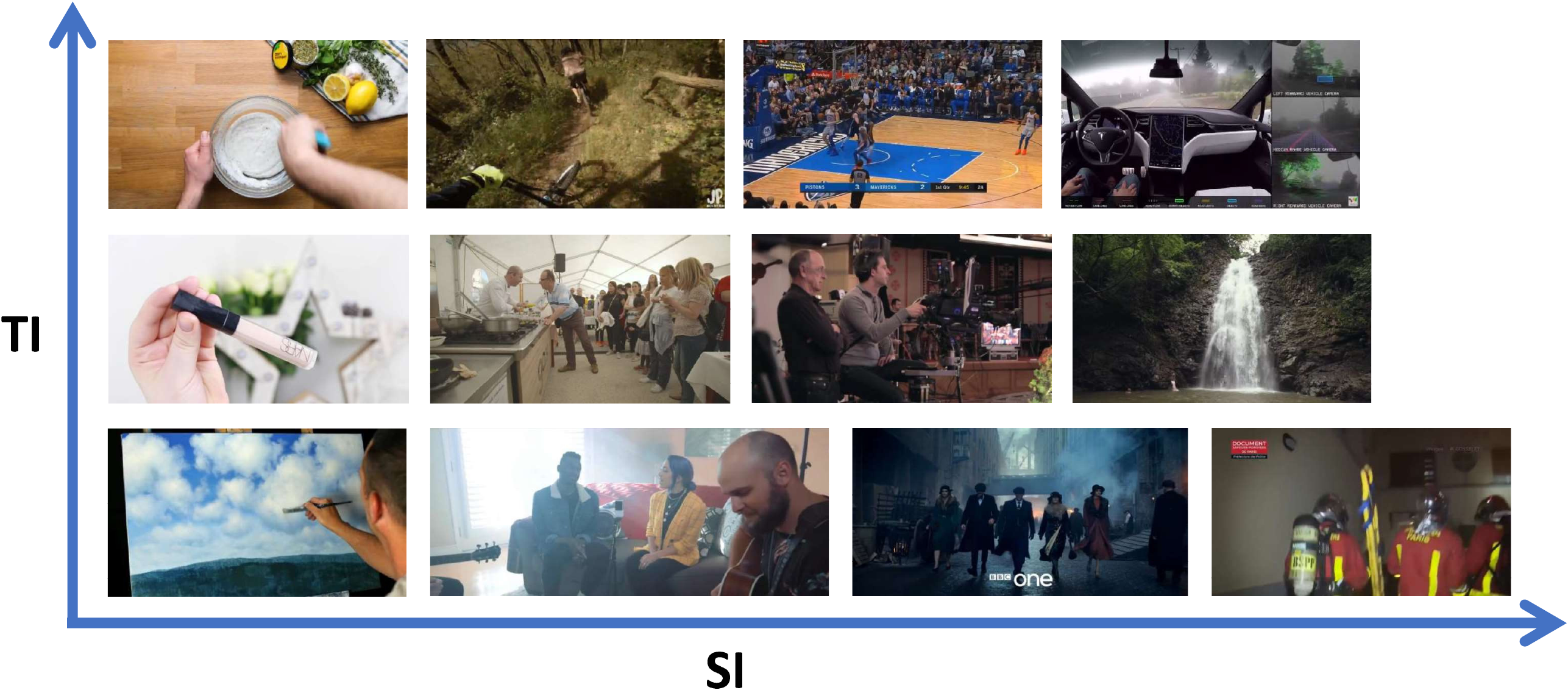}
  \caption{Thumbnails of the 12 videos watched by users in the first study. Thumbnails are ordered along the SI and TI dimensions.}
  \label{fig:study1_thumbnails}
\end{figure}

The volunteers in the first study were 22 students from our institution with both technical and non-technical backgrounds. The group consisted of 13 male and 9 female participants. We select 12 one-minute-long YouTube videos to be watched by the users (a preview of these videos is shown in Figure~\ref{fig:study1_thumbnails}). The video content varied among the videos from music, sports, outdoor/indoor activities, and others, resulting in various spatial and temporal characteristics of the videos. We computed the average SI and TI for all 12 videos, and the results are shown in Table~\ref{tab:siti1}. These numbers illustrate the heterogeneity in the video content with regard to their spatial and temporal features. 

\begin{table}
\centering
\caption{Spatial information (SI) and Temporal information (TI) indices for the videos used in the first user study.}
\label{tab:siti1}
\begin{tabular}{lcc}
   \toprule
Video ID & Average SI & Average TI  \\ 
 \midrule
1                 & 55.51               & 19.45                \\ 
2                 & 117.26              & 26.58              \\ 
3                 & 52.59               & 7.77               \\ 
4                 & 61.69               & 15.39               \\ 
5                 & 59.32               & 29.42               \\ 
6                 & 29.05               & 11.52              \\ 
7                 & 56.65               & 9.72                 \\ 
8                 & 46.14               & 8.81               \\ 
9                 & 39.77               & 11.41               \\ 
10                & 80.04               & 19.03                 \\ 
11                & 126.88              & 13.85               \\ 
12                & 36.38               & 8.60                 \\ 
  \bottomrule
\end{tabular}
\end{table}

Each of the participants in the study group watched videos in different activity states (three videos per state): still, walking, running, and traveling as a passenger in a vehicle. All the experiments were performed on the campus of Faculty of Computer and Information Science in Ljubljana, Slovenia: in the same laboratory room when still, on the same hallway when walking and running, and on the same route on the campus when traveling as a passenger in a vehicle (the same driver and vehicle for all tests/subjects). The following smartphones were used during this study for watching videos by the participants: Samsung Galaxy S3, Samsung Galaxy S4 and Nexus 6.

To ensure the obtained results were comparable and relevant, all participants were instructed to follow the same protocol during the experiments. Hence, the following instructions were given to the participants:
\begin{itemize}
    \item The users were instructed about the resolutions available and the process of changing the resolution when watching a video. They were asked to switch the resolution to a higher one only when dissatisfied with the quality;
    \item They were asked to keep the device horizontal at all times to ensure the video is played in full-screen;
    \item Users were allowed to change sound volume and use headphones during the experiments according to their preferences;
    \item The brightness was pre-set to 80\% and the participants were asked not to change it;
    \item Before each experiment the users were informed about the video and the resolution they should start the experiment with; the starting resolutions presented a pseudorandom distribution. We choose this approach to avoid the situation where always starting from a low resolution might artificially reduce the inferred viewer's expectations, as viewers might be inclined to proceed with the default resolution. 
\end{itemize}

\subsection{User Study 2: Video Properties and User Personality vs Video Resolution Requirements}

\begin{figure}
\centering
  \includegraphics[scale=0.5]{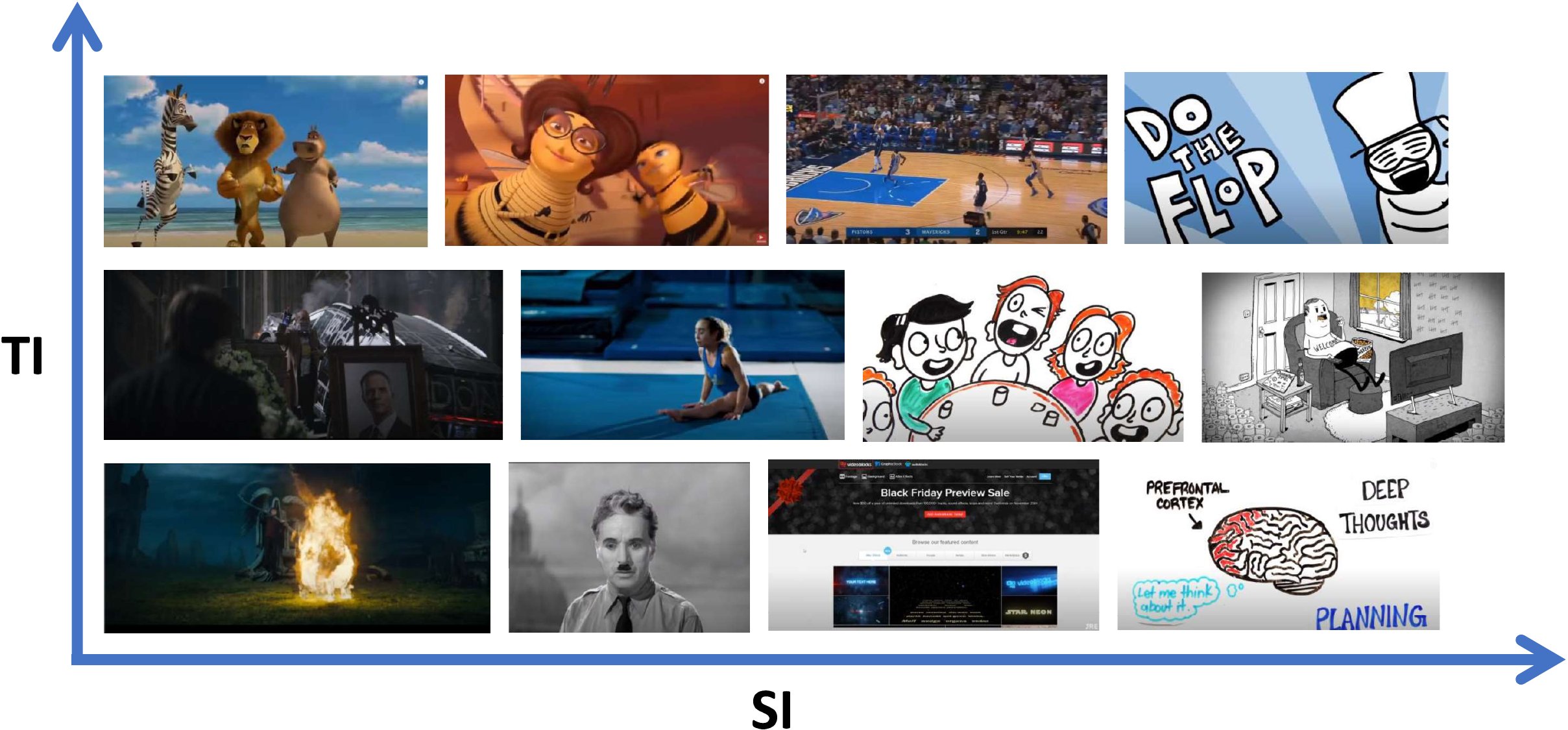}
  \caption{Thumbnails of the 12 videos watched by users in the second study. Thumbnails are ordered along the SI and TI dimensions.}
  \label{fig:study2_thumbnails}
\end{figure}

We conducted the second study with 23 users, 
13 male and 10 female. Each user watched 4 videos in each of the following mobility states: still, walking and running. Due to the COVID-19 pandemic restrictions in place at the time of this second study, having a researcher driving a car with participants was unfeasible, as such this mobility state was not recorded. To examine how the spatial and temporal complexity of the videos impact the user's quality expectations in the mobility states, the videos were selected so that their SI/TI scores fall in the following categories: low SI \& low TI, low SI \& high TI, high SI \& low TI and high SI \& high TI. While a review of related scientific literature revealed no "absolute" scale for SI and TI metrics, based on the results of the first study (in terms of SI/TI values for which the highest correlations were observed) and other related work~\cite{8858486}, for the purpose of this study we considered the following thresholds: Low SI $<=40$, High SI $>=110$, Low TI $<=10$, High TI $>=25$.
Consequently, a total of 12 1-minute long videos were selected, with 3 videos in each of the aforementioned SI/TI categories. A thumbnail preview of the videos in this second study can be seen in Figure \ref{fig:study2_thumbnails}.

\begin{table}
\centering
\caption{Spatial information (SI) and Temporal information (TI) indices for the videos used in the second user study, and their corresponding grouping into categories.}
\label{tab:siti1}
\begin{tabular}{llll}
\toprule
Video ID & Category            & \multicolumn{1}{c}{Average SI} & \multicolumn{1}{c}{Average TI} \\ \midrule
1        & Low SI, Low TI   & 8.77                          & 3.68                          \\
2        & Low SI, High TI  & 38.86                         & 36.95                         \\
3        & High SI, Low TI  & 138.69                        & 8.30                          \\
4        & High SI, High TI & 136.49                        & 28.59                          \\
5        & Low SI, Low TI   & 14.34                         & 3.49                          \\
6        & Low SI, High TI  & 40.17                         & 38.44                         \\
7        & High SI, Low TI  & 117.56                        & 6.14                          \\
8        & High SI, High TI & 118.66                        & 26.81                         \\
9        & Low SI, Low TI   & 41.16                         & 1.42                          \\
10       & Low SI, High TI  & 40.30                         & 31.63                         \\
11       & High SI, Low TI  & 123.35                        & 8.14                          \\
12       & High SI, High TI & 138.75                        & 37.13                         \\  \bottomrule
\end{tabular}
\end{table}

The experiments were performed on the personal smartphones of the users at the same locations in Rosenheim, Germany. Given that this study was performed under the COVID-19 pandemic restrictions, semi-outdoor spaces were used: a personal garage for the still experiments, and a public parking (Parkhaus P12 Bahnhof Nord) for the walking and running experiments.

Again, all participants were instructed to follow the same protocol as the first study (described above) during the experiments. In addition, the following specific issues were addressed: 
\begin{itemize}
\item In this study, the following resolutions were available: 360p, 480p, 720p and 1080p. The lowest resolutions available in the first study were discarded because it was shown they had no significant impact on both the final resolution users ended up watching the videos in and the energy consumption (the lowest three resolutions: 144p, 240p and 360p have very similar energy consumption);
\item In light of the above, and also since we noticed from the first study that viewers are not reluctant to change the initial resolution, the starting resolution in this study was always the lowest one (i.e. 360p);
\item The users performed the activities in a cyclic order so that every consecutive user performs the activities in a different order when compared to the previous user. (e.g. User $n$: still, walking, running; User $n+1$: running, still, walking, etc.); 
\item The same cyclic approach was used for the types of videos that users watched while in each of the mobility states, e.g. User $n$ still: video 1 (low SI \& low TI), video 2 (low SI \& high TI), video 3(high SI \& low TI), video 4(high SI \& high TI); User $n+1$ still: video 1 (high SI \& high TI), video 2 (low SI \& low TI), video 3 (low SI \& high TI), video 4 (high SI \& low TI), etc. We employed this ordering of activities and video categories to minimize the overlap of activity-video category items over users;
\item In addition to the demographics data (age, gender), the smartphone model used in the experiments and whether or not the user had glasses, we also collected information on a user's personality by administering the 10-item short version of the Big Five Inventory (the BFI-10 test)~\cite{rammstedt2007measuring}. 
\end{itemize}

\section{Results}
\label{sec:results}

Based on the conducted user studies, in this section we examine how the viewer's satisfaction and quality expectations are impacted by the physical activity by analyzing the resolutions that were found acceptable when watching videos in each of the four mobility states. Next, we perform a statistical investigation to determine how the video content (its spatial and temporal characteristics) impacts the viewer's tolerance to lower video quality. Aside from the viewer context and the video content, viewer-related factors are also shown to play a role. As such, we also address the impact that viewer's personality traits have on the required video quality by using hierarchical modelling (performing mixed effects modelling using personality as a random effect grouping factor). Finally, based on these three dimensions, we analyse the suitability of predictive mobile video resolution models. 

\subsection{The Role of Physical Activity}
\label{sec:results:physical}

\begin{figure}
\centering
\subfigure[Study 1]{\label{fig:activity:study1}\includegraphics[scale=0.55]{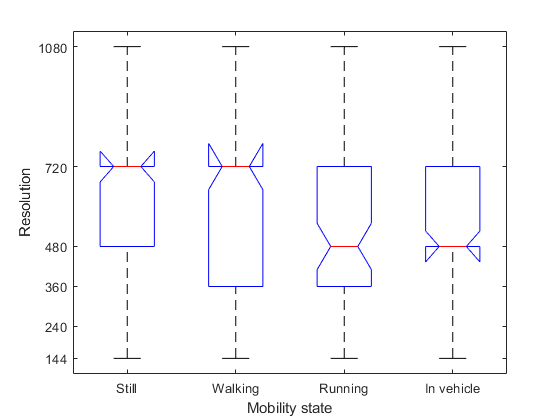}}
\subfigure[Study 2] {\label{fig:activity:study2}\includegraphics[scale=0.55]{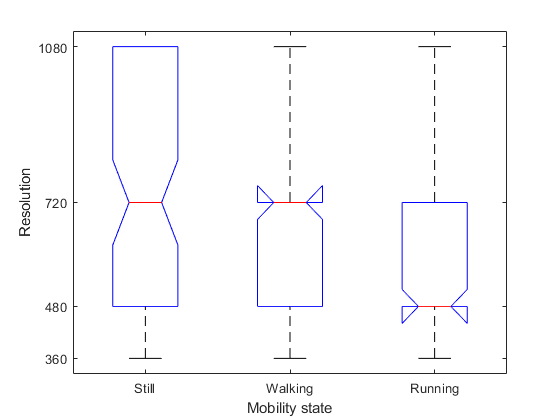}}
\caption{Boxplot depiction of the distribution of resolutions in which viewers completed watching videos in each activity state in each of the two studies. Central line in each box: median; edges of the boxes: 25th and 75th percentiles of the distributions; Whiskers: most extreme data points not considered outliers.}
\label{fig:activity}
\end{figure}

To illustrate the role of the physical activity of the viewer on the resolution, we plot the distribution of the final resolutions in which viewers completed watching videos while in each of the activity states in both studies in Figure~\ref{fig:activity}.

The results, which are consistent for both studies, are in favor of the \emph{RQ\textsubscript{1}} hypothesis that the activity context of the viewer impacts the perception of the video quality, and ultimately the satisfaction with the viewing experience. Thus, the data shows viewers are satisfied with higher resolutions when they watch the video while still (the median of the distribution is highest for this activity, at 720p). This is expected, since in such situations a viewer can fully concentrate on the video. The next highest average resolution is found in case the viewers are walking. In this state the distribution tails are more prominent in the first study, and while the median of distribution remains as high as it was with viewers being still in both studies (i.e. 720p), the $25^{th}$-percentile of distribution in the first study is at 360p (c.f. 480p for still viewers). 

Riding as a passenger in a vehicle induces further tolerance towards lower resolutions, with the median of the acceptable resolution dropping to 480p, yet the distribution becomes more ``compact" than it is the case with the distribution observed when the viewers are in the walking state. We suspect that the effect stems from varying abilities of our viewers to simultaneously walk and pay attention to the video. For some such multitasking may be a routine endeavor, thus, they require a higher resolution, whereas others might find it difficult to pay attention to the videos and regard the resolution unimportant.

Finally, the running state leads to a further drop of resolution distribution, with the the $25^{th}$-percentile at 360p and the median at 480p. This is not surprising since when engaged in a intense physical activity the viewer is less likely to be focused on the screen for anything but brief periods of time. By having to divide the attention between the video and the surroundings, the viewers find lower resolutions acceptable since they do not have the time to notice imperfectly rendered details.

\begin{figure}
\centering
\subfigure[Study 1]{\label{fig:time_before_change:study1}\includegraphics[scale=0.57]{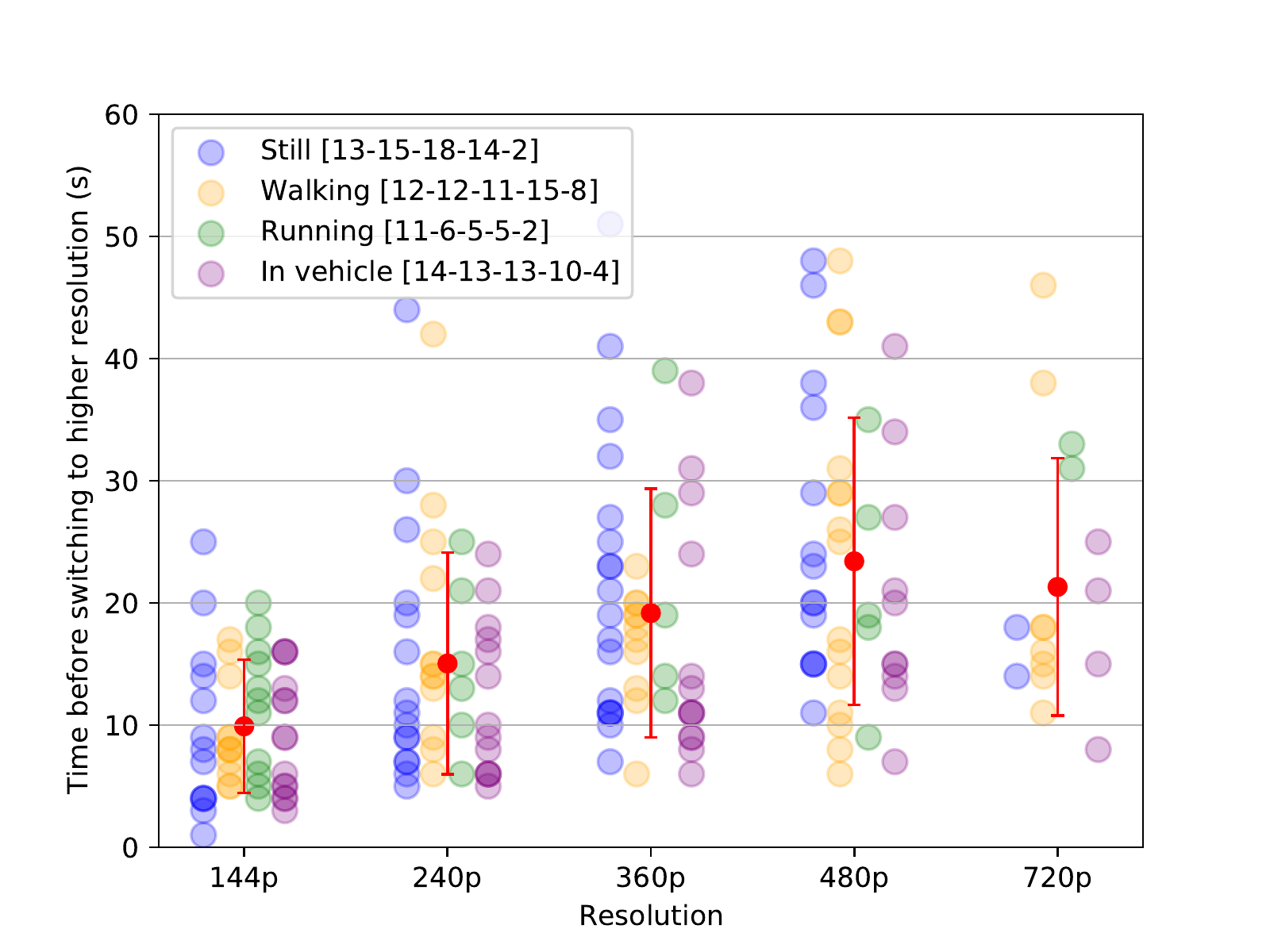}}
\subfigure[Study 2] {\label{fig:time_before_change:study2}\includegraphics[scale=0.36]{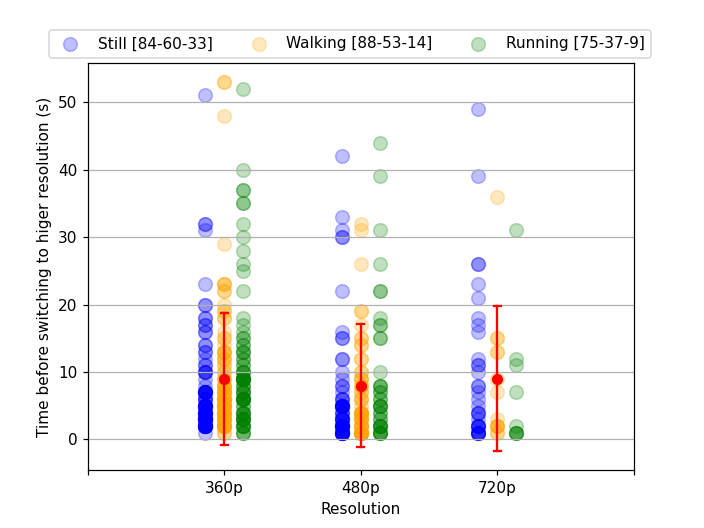}}
\caption{Time elapsed before viewers switching to a higher resolution for different activity states. A colored circle marks the moment in time the viewer increased resolution while watching the video. The red dot is the average represented with relation to two standard deviations (the red segment's extremities). In the legend, the values indicate the number of changes performed by the viewers in each of the resolutions.}
\label{fig:time_before_change}
\end{figure}

To help understand viewer behavior in each activity state, Figure~\ref{fig:time_before_change} shows all the changes in resolution performed by the viewers in the four activity states and the time elapsed before each change was made. In the legend the number of changes in each resolution for each mobility state can be observed. These results confirm that viewers had the lowest quality expectations (or highest tolerance to lower quality) while running, since in this state they made the lowest number of switches to higher resolution (the green circles on the chart). The highest number of instances where the viewers switched to higher resolutions can be observed in the still state, confirming that when in this activity state, viewers have the highest quality expectations. Finally, irrespective of the physical activity, as we move from the lower resolutions to the higher ones we observe a slight increase in the time to switch to a higher resolution, which confirms that the viewers complied with the protocol, i.e. switched the resolution only when not satisfied with the current one.

We then performed the statistical analysis of the results for both studies. A Kruskal-Wallis test shows that there is indeed a significant difference in the acceptable resolution depending on the activity state: $H(3) = 14.139, p<0.003$ for the first study, $H(2) = 19.817, p<0.001$ for the second. This confirms the hypothesis that the activity state influences the viewer's video quality requirements. To assess the strength of the relationship between the context and the resolution we computed the effect size estimate for the Kruskal-Wallis result \cite{tomczak2014need}.
More specifically, we computed the eta-squared measure ($\eta ^{2}$) using the following formula \cite{cohen2008explaining}:

\begin{equation}
    \eta ^{2}_{H}=\frac{H-k+1}{n-k}
    \label{eq:etasquared}
\end{equation}
where $H$ is the Kruskal-Wallis H-test statistic, $k$ is the number of groups and $n$ the total number of observations.

Eta-squared estimate assumes values from 0 to 1 and multiplied by 100 indicates the percentage of variance in the dependent variable explained by the independent variable \cite{tomczak2014need}. For our experiment the computed eta-squared was $0.04$ for the first study and $0.06$ for the second study; in the related scientific literature \cite{cohen2008explaining} eta-squared values less than $0.06$ account for a small (weak) effect. Thus, while there is a statistically significant relationship between the activity state and the resolution, this relationship is shown to be weak.

\subsection{The Role of Spatial and Temporal Properties of a Video}
\label{sec:results:properties}

In light of the above statistical results, which indicate that other factors might influence a viewer's satisfaction with lower resolutions in different mobility states, we analyzed the impact of the video content on a viewer's receptivity to different video resolutions. The Kruskal-Wallis test shows that there is a statistically significant relationship between \textit{the actual video content being played} and viewer's quality expectations (resolution found acceptable): $H(11) = 65.328, p<0.001$ for the first study, $H(11) = 79.045, p<0.001$ for the second. For evaluating the strength of this relationship we computed the same eta-squared effect size measure using Equation~\ref{eq:etasquared}, with the results for the two studies being $0.20$ and $0.25$, respectively. Based on the related literature \cite{cohen2008explaining}, values higher than $0.14$ indicate a large effect. This confirms \emph{RQ\textsubscript{2}}, i.e. that there is a strong relationship between the video content and the viewer's quality expectations when watching the video in specific mobility states.

A surprisingly strong effect of the individual video content warrants further investigation of particular aspects of a video that influence a viewer's decision to require a higher playback resolution. A viewer's perception of the content can stem from the visual elements depicted in the video, speed of scene changes, colours, and other technical elements, but could also stem from the relationship between the viewer and the video content, including the viewer's interest in a particular topic, previous exposure to that and similar videos, to name a few factors. In this work, however, we aim to uncover factors that could be easily harnessed for automatic playback resolution adaptation. Thus, we focus on the spatial (SI) and temporal (TI) complexity indices readily obtainable from a downloaded video.

To evaluate how the spatial and temporal complexity of the videos relates to the viewer quality perception of the videos in each mobility state we analyzed the link between the average resolution of the videos viewed in each state versus their SI and TI scores. We computed the Pearson correlation coefficient between the resolution and average SI and TI values for each mobility states, and the results are shown in Table~\ref{tab:pearson}.

The strongest link between the selected playback resolution and the SI is observed when a viewer is running (a Pearson correlation of $0.86$). Running is of a particular interest to this study since it is the mobility state where one would expect the viewer’s satisfaction requirements to drop the most. This strong link shows that when a viewer is physically active (e.g. running), the required video quality and the spatial complexity of the video being played exhibit a strong positive linear correlation (i.e. the higher the spatial complexity of the video, the higher the required resolution). Out of the videos watched by the viewers while running in the first study, for videos $10$ and $11$ that have the highest SI scores, the viewers required the highest resolutions.

\begin{table}
\centering
\caption{Pearson correlation coefficient between the final selected resolution and the average video SI/TI when a viewer is in a particular mobility state in the first user study.}
\label{tab:pearson}
\begin{tabular}{lcc}
\toprule
           & Resolution vs. SI & Resolution vs. TI \\
\midrule
Still      & -0.05             & 0.21              \\
Walking    & 0.31              & 0.54              \\
Running    & 0.86              & 0.23              \\
In vehicle & -0.28             & -0.17             \\
\bottomrule
\end{tabular}
\end{table}



With regard to the link between the average resolution of all videos watched by all viewers in each mobility state and their corresponding TI score, the Pearson correlation analysis indicates that a moderate positive linear correlation is present when the viewer is in mobility states requiring moderate physical movement, such as walking, where the coefficient is $0.54$. While walking the viewers requested the highest average resolution for video number $5$, which has the highest TI score among the videos watched while walking.



\begin{figure}
\centering
  \includegraphics[width=0.8\columnwidth]{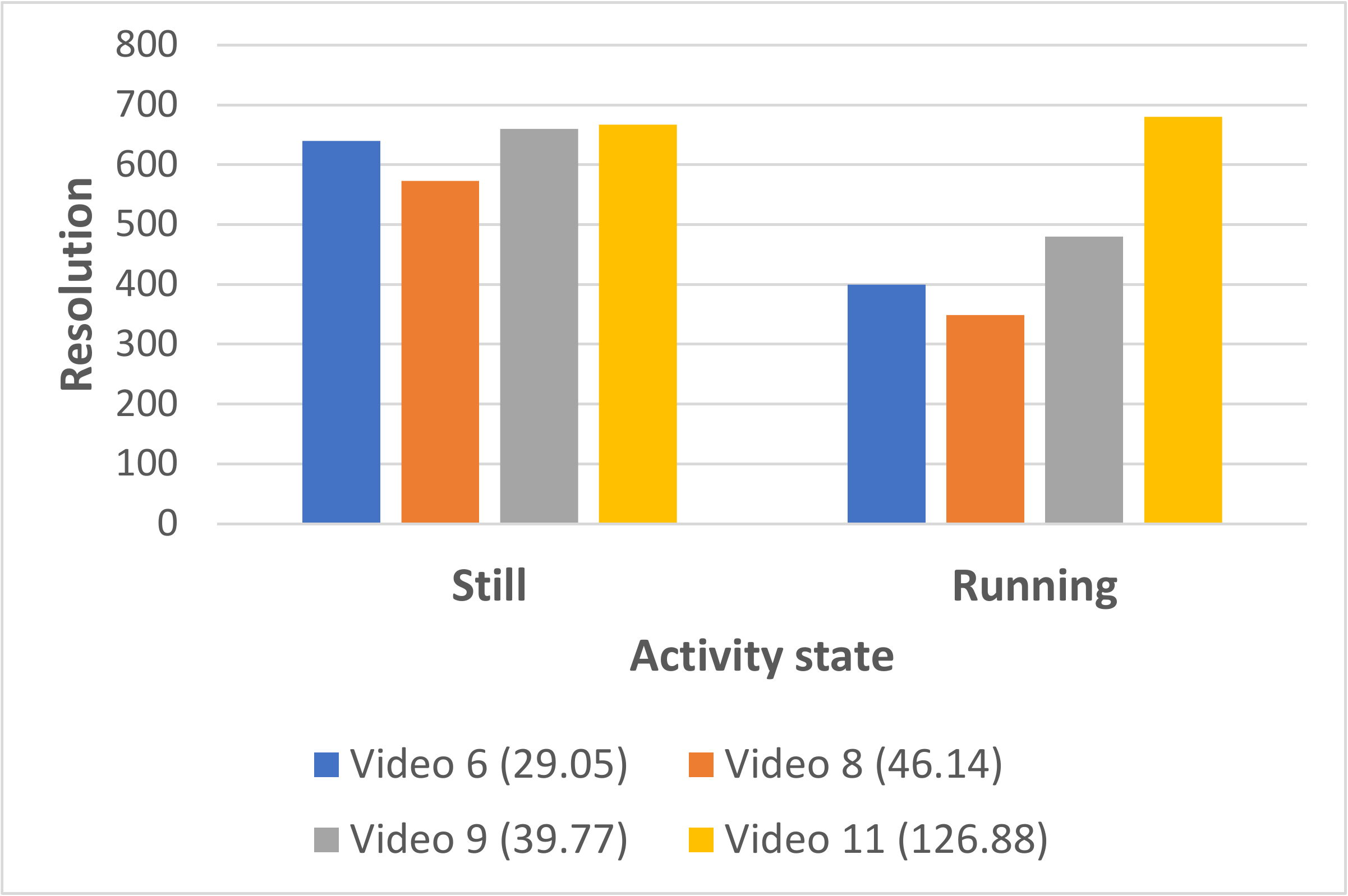}
  \caption{Average resolution in still vs. running for selected videos (and their corresponding SI values) in the first user study. While the general trend is that a running viewer is satisfied with a lower resolution than a walking viewer, a high SI video (Video 11) leads to higher resolution requirements when a viewer is running.}
  \label{fig:still_vs_running}
\end{figure}

To better illustrate how the spatial and temporal characteristics of a video influence the viewer's quality perception in different activity states, Figure~\ref{fig:still_vs_running} shows how a selection of videos are perceived by the viewers when standing still vs. running (a subset comprising all videos which viewers watched in both activity states: videos $6$, $8$, $9$ and $11$). The plot displays the average resolution for each video in each of the two activity states, and it is noticeable that videos $6$, $8$ and $9$ show a similar behavior, i.e. they score similar average resolutions when still (between $650$ and $550p$) and their average resolutions drop considerably during running (between $350$ and $500p$). Video $11$ however has a different behavior: while it also has an average resolution of about $650p$ while standing still, it does not decrease while running, on the contrary it slightly increases. The reason behind this phenomenon is that video $11$ has the highest spatial information index among all 12 videos, and thus viewers perceptually require higher resolutions when running and viewing this video, compared to the other videos with lower spatial complexity.

To statistically examine the interplay between the physical activity and the video content and its role on a viewer's expectations we created a linear regression model where the dependent variable is the resolution and the explanatory variables are the activity states, SI, TI, and the cross-products representing the interaction effects between the activity states and the SI/TI scores. The results of this linear regression are presented in Table~\ref{tab:lrm}.

\begin{table}
\centering
\caption{Linear regression results for the resolution as the dependent variable - results from the first study.}
\label{tab:lrm}
\begin{tabular}{lcc}
   \toprule
Variable         & Coefficient & p-value         \\
 \midrule
Intercept            & 647.37   & \textless 2e-16\textsuperscript{***} \\
walking              & -247.89  & 0.02\textsuperscript{**}            \\
running              & -394.62  & \textless 0.01\textsuperscript{***}  \\
in\_vehicle          & -65.83   & 0.48            \\
spatial              & 0.31     & 0.73            \\
temporal             & -3.72    & 0.41            \\
walking:spatial      & 0.25     & 0.87            \\
running:spatial      & 2.93     & 0.05\textsuperscript{*}            \\
in\_vehicle:spatial  & -1.16    & 0.37            \\
walking:temporal     & 13.15    & 0.05\textsuperscript{*}            \\
running:temporal     & 5.53     & 0.45            \\
in\_vehicle:temporal & 6.03     & 0.30            \\
 \midrule
\multicolumn{3}{l}{Multiple R-squared: 0.1094}    \\
\multicolumn{3}{l}{Adjusted R-squared: 0.0705}  \\
\multicolumn{3}{l}{Standard error of the estimate: 218.3}\\
\bottomrule
\addlinespace[1ex]
\multicolumn{3}{l}{\textsuperscript{***}$p<0.01$, 
  \textsuperscript{**}$p<0.05$, 
  \textsuperscript{*}$p<0.1$}
\end{tabular}
\end{table}

The regression shows the impact of a particular activity and the specific spatial and temporal complexity of a video on the required resolution. When viewers are walking or running, they require a lower resolution as indicated by the strong negative coefficients and low p-values; the effect is less pronounced when in-vehicle. The effects of the spatial and temporal complexity of a video on the required resolutions are not relevant by themselves (non-significant values for "temporal" and "spatial"), only in interaction with certain activities. As such, high temporal information videos require higher resolution when a viewer is walking (as indicated by the low p-value of $0.05$ and thus confirming the correlation illustrated in Table~\ref{tab:pearson}). In addition, higher spatial information videos require higher resolutions when a viewer is running (the low p-value of $0.05$ confirms the correlation also illustrated in Table~\ref{tab:pearson}).

In addition to the above, however, the linear regression R-squared value is low, indicating that the model does not fully explain the data. This may stem from the limited data collected in our user study. More specifically, not all videos where watched in all activity states and not all videos were watched by all viewers. Furthermore, low R-squared value is likely an indicator that other contextual variables not considered in our study (e.g. outside noise, a viewer's interest in the video content, etc.) may impact quality expectations.

\subsection{The Role of Personality}
\label{sec:results:personality}

The exploratory analysis conducted on data collected during our first study demonstrates that both the context in which a video is watched as well as the content of the video play a role in the final playback resolution that a viewer is satisfied with. Yet, our first study does not allow further analysis of the role of individual user's traits on the watching behaviour. 

In the second study we collected information about our participants' personalities using the BFI-10 test. For investigating the role of personality on the required resolution, we performed the Kruskal-Wallis test and uncovered a statistically significant relationship between \textit{the dominant personality of a viewer} and his/hers quality expectations (resolution found acceptable): $H(4) = 15.874, p<0.003$. The eta-squared effect size (Equation~\ref{eq:etasquared}) amounts to $0.04$ indicating a weak effect \cite{cohen2008explaining}. This confirms \emph{RQ\textsubscript{3}}, i.e. that the viewer's personality traits impact the quality requirements in terms of playback resolution when watching a video on a mobile device. However, the effect size shows this impact to be weak.

We next create two linear regression models to additionally explore the statistical interplay between the personality and the end resolution required by the viewers in the second user study. To ensure that the personality does not ``hide'' other factors, we explicitly include the demographics as well. In the first model we encoded the dominant personality trait of each user as a variable. The regression confirmed that personality plays a significant role in the end resolution required by a viewer. The detailed results of this linear regression are presented in Table~\ref{tab:lrm_pers1}.

\begin{table}
\centering
\caption{Linear regression results for the resolution as the dependent variable - results from the second study with dominant personality trait as a variable.}
\label{tab:lrm_pers1}
\begin{tabular}{lcc}
   \toprule
Variable         & Coefficient & p-value         \\
 \midrule
Intercept            & 1033   & \textless 0.001\textsuperscript{***} \\
Activity             & -79.382  & \textless 0.001\textsuperscript{***} \\
SI                   & -1.30    & \textless 0.001\textsuperscript{***} \\
TI                   & -0.61    & 0.49            \\
Gender               & 25.69   & 0.36            \\
Age                  & 1.06    & 0.6 \\
Glasses              & 21.2     & 0.48 \\
Personality          & -42.896 & \textless 0.001\textsuperscript{***}  \\
 \midrule
\multicolumn{3}{l}{Multiple R-squared: 0.228}    \\
\multicolumn{3}{l}{Adjusted R-squared: 0.2}  \\
\multicolumn{3}{l}{Standard error of the estimate: 214.2}  \\
\bottomrule
\addlinespace[1ex]
\multicolumn{3}{l}{\textsuperscript{***}$p<0.01$, 
  \textsuperscript{**}$p<0.05$, 
  \textsuperscript{*}$p<0.1$}
\end{tabular}
\end{table}

To investigate the effect that each particular dominant personality type has on the end resolution, the second model encoded the distinct personality traits percentiles as variables. The results of this regression model (illustrated in Table~\ref{tab:lrm_pers2}), show that of the five dominant personality traits, three are shown to have a significant influence on the end resolution: agreeableness, conscientiousness and neuroticism all correlate with higher end resolutions. Openness is the only dominant personality trait that correlates with a lower resolutions, but this dependency is not shown to be statistically significant.

\begin{table}
\centering
\caption{Linear regression results for the resolution as the dependent variable - results from the second study with distinct personality traits percentiles as variables.}
\label{tab:lrm_pers2}
\begin{tabular}{lcc}
   \toprule
Variable         & Coefficient & p-value         \\
 \midrule
Intercept            & 657.25   & \textless 0.001\textsuperscript{***} \\
Activity             & -79.382  & \textless 0.001\textsuperscript{***} \\
SI                   & -1.30    & \textless 0.001\textsuperscript{***} \\
TI                   & -0.61    & 0.49            \\
Gender               & -3.079   & 0.91            \\
Age                  & 2.871    & 0.17 \\
Glasses              & 82.1     & 0.01\textsuperscript{**}           \\
Extraversion         & 9.37     & 0.82            \\
Agreeableness        & 139.28   & 0.004\textsuperscript{***}            \\
Openness             & -18.438  & 0.74            \\
Conscientiousness    & 160.792  & 0.002\textsuperscript{***}            \\
Neuroticism          & 96.871   & 0.037\textsuperscript{**} \\
 \midrule
\multicolumn{3}{l}{Multiple R-squared: 0.232}    \\
\multicolumn{3}{l}{Adjusted R-squared: 0.2}  \\
\multicolumn{3}{l}{Standard error of the estimate: 215.3}  \\
\bottomrule
\addlinespace[1ex]
\multicolumn{3}{l}{\textsuperscript{***}$p<0.01$, 
  \textsuperscript{**}$p<0.05$, 
  \textsuperscript{*}$p<0.1$}
\end{tabular}
\end{table}

\begin{figure}
\centering
  \includegraphics[scale=0.7]{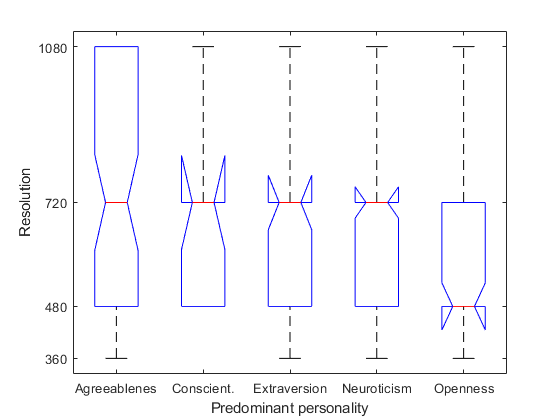}
  \caption{Boxplot depiction of the distribution of resolutions in which the viewers completed watching videos and their dominant personality trait - data from the second study.}
  \label{fig:boxplot_personality}
\end{figure}

\subsection{Hierarchical Modelling}
\label{sec:results:hierarchical}

Concluding that the first three of our research hypotheses hold, i.e. that a viewer's physical activity at the time of watching the video, the video's content, and the viewer's personality all impact the desired mobile video playback resolution, we now proceed with modelling the joint impact of these factors. 

Mixed-effect modelling represents a statistical instrument primarily used to describe relationships between a response variable and some covariates in data that are grouped according to one or more classification factors. A mixed effects model has both fixed effects (parameters associated with an entire population) and random effects (which are associated with individual experimental units drawn at random from a population)~\cite{Pinheiro2000}.

For building these models we adopt an incremental, iterative approach in which we gradually increase the complexity of the previously built model by adding an additional parameter, either as a fixed or as a random effect. To guide our approach and evaluate the appropriateness of each model, we use AIC (Akaike information criterion), BIC (Bayesian information criterion) and the R-squared measure marginal vs. conditional (expressed by fixed effects vs. both fixed and random effects). AIC and BIC are the two most commonly used penalized model selection criteria~\cite{kuha2004aic}. AIC penalizes the inclusion of additional variables to a model. It adds a penalty that increases the error when including additional terms. As such, a lower AIC score is an indicator of a better model. BIC is a variant of AIC with a stronger penalty for including additional variables to the model~\cite{kassambara2018machine}.

We run a mixed-effects model analysis using R with the lme4 package, and start from an intercept-only model that allows evaluating the appropriateness of the grouping variable – dominant personality trait. For this model, the regression function intercept varies across different personality types. We calculate the intraclass correlation coefficient (ICC) to get an estimate of how much of the end resolution variation is explained by clustering along the dominant personality, and obtain a score of 0.03, indicating a weak grouping. 

We then move on to add fixed effects parameters, and we incrementally add Activity, SI and TI. When adding Activity, and further on SI, both AIC and BIC scores decrease, however after adding TI they both increase. In addition R-squared scores (both marginal and conditional values, computed with the $squaredGLMM$ function) increase after adding Activity, and even more after adding SI, but stay constant after adding TI. As such we drop TI as a fixed effects parameter. Inspecting this latest model we notice that when viewers are still they require much higher resolutions compared to when engaged in the other two mobility states. We next add the interaction between Activity and SI, which improves the model even further. Using Gender as part of the fixed effects parameters does not improve the model, both with regard to the AIC and BIC scores, and R-squared values. However, accounting for the interaction between Gender and SI is shown to improve the model. We notice that male viewers require lower end resolutions than female viewers only for videos with lower SI scores, while for videos with high SI this trend is reversed. Finally, by adding Glasses as fixed effects term, the model slightly improves further, and it illustrates that viewers wearing glasses tend to require higher end resolutions as the SI of the video increases, compared to viewers not wearing glasses. Adding the last remaining parameter, age, to fixed effects, does not improve the model.

To conclude, the final best mixed effects model includes dominant personality trait as a random effect (grouping factor), and the following parameters as fixed effects: Activity, SI, Gender, Glasses, with the interaction variables between Activity and SI, and Gender and SI, respectively. Table~\ref{tab:mixed} shows the detailed results of the mixed effect model analysis. 

\begin{table}
\centering
\caption{Mixed effects model on the second study data for the end resolution as the dependent variable, personality as the grouping factor, and SI, Activity, Gender, SI*Activity, SI*Gender as fixed effects.}
\label{tab:mixed}
\begin{tabular}{lcc}
   \toprule
Random effects:         &  &          \\
Groups         & Variance & Std.Dev.         \\
 \midrule
Personality (Intercept)            & 3612   & 60.1 \\
Residual             & 43677  & 209.0 \\
   \toprule
Fixed effects:         &  &          \\
Name         & Estimate & p-value         \\
 \midrule
Intercept            & 750.32   & \textless 0.001\textsuperscript{***} \\
SI             & -2.27  & \textless 0.001\textsuperscript{***} \\
ActivityStill                   & 222.70    & \textless 0.001\textsuperscript{***} \\
ActivityWalking                   & -17.08    & 0.78            \\
GenderMale               & -103.89   & 0.03\textsuperscript{**}            \\
Glasses & 9.55 & 0.85 \\
SI:ActivityStill                  & -0.79    & 0.19 \\
SI:ActivityWalking                  & 1.07    & 0.11 \\
SI:GenderMale                  & 1.30    & 0.01\textsuperscript{**} \\
SI:Glasses & 0.46 & 0.36 \\
 \midrule
\multicolumn{3}{l}{R2m: 0.21}    \\
\multicolumn{3}{l}{R2c: 0.27}  \\
\bottomrule
\addlinespace[1ex]
\multicolumn{3}{l}{\textsuperscript{***}$p<0.01$, 
  \textsuperscript{**}$p<0.05$, 
  \textsuperscript{*}$p<0.1$}
\end{tabular}
\end{table}

The random effects analysis of the model shows that the differences between different dominant personality types explain just $\sim 8 \%$ (3612) out of the total variance (3612 + 43677) ``left over'' after the variance explained by the fixed effects. We analyzed the amount of variance explained by fixed vs. random factors via the $r.squaredGLMM$ function computing pseudo R2 for mixed-models. We obtained R2 = 0.21 for the variance explained by fixed factors, and R2 = 0.27 for the variance explained by both fixed and random factors, showing that the differences in personality types explain approximately 22\% of the total variance explained by our model.

The fixed effects analysis of the model shows that viewers require a higher resolution when still (the estimate value for still is $222.7$ vs. $-17.08$ for walking), and that videos with a high SI require slightly lower resolutions (SI estimate is $-2.27$). However, male viewers are shown to require slightly higher resolutions as the videos have higher SI (GenderMale has an estimate of $-103.89$, while SI:GenderMale has a positive estimate of $1.30$). 

\begin{figure}
  \includegraphics[width=\columnwidth]{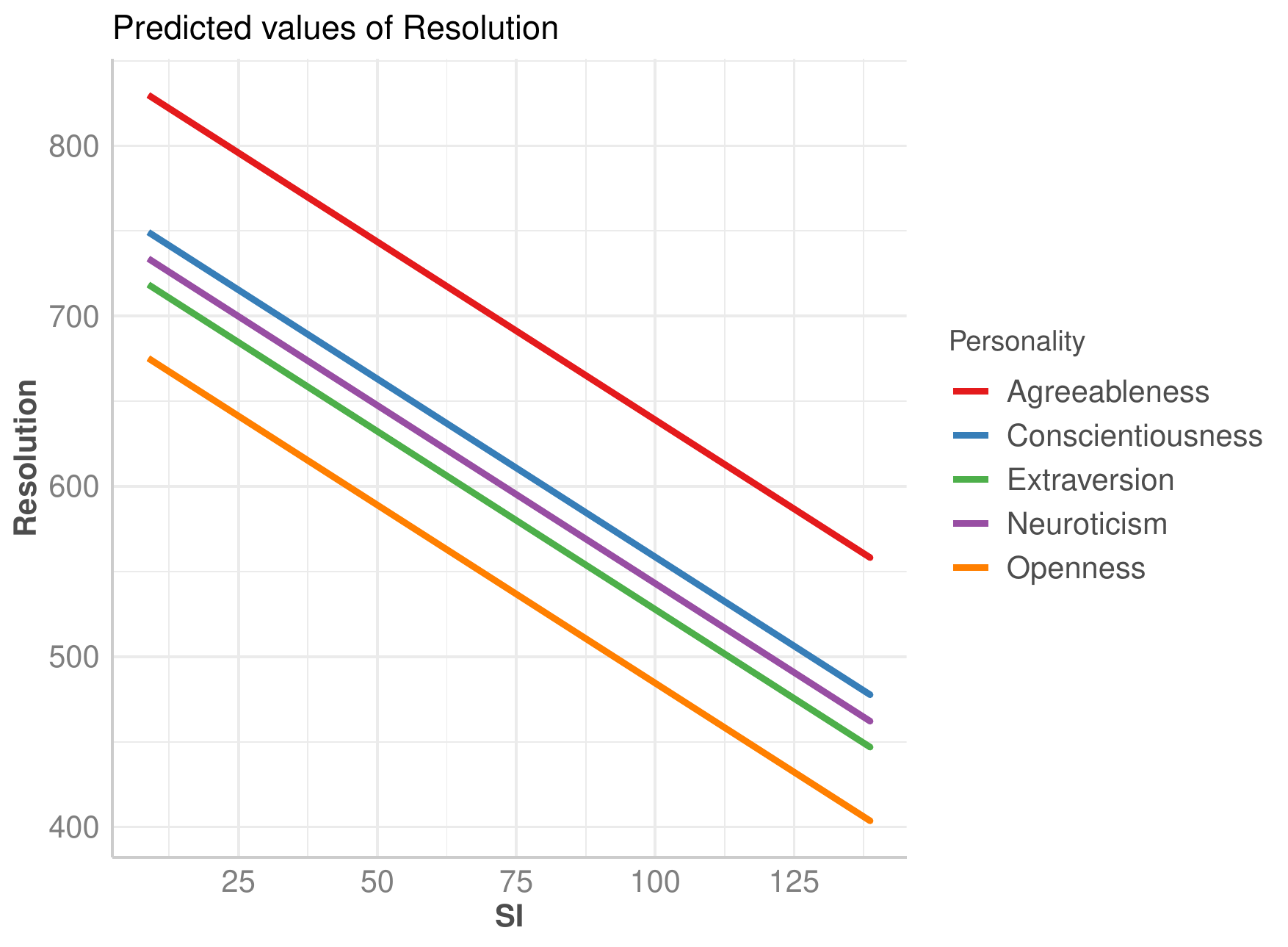}
  \caption{Desired resolution vs. SI for each dominant personality trait. Agreeableness stands out as requiring the highest resolution while openness requires the lowest.}
  \label{fig:endres_vs_si}
\end{figure}

Figure~\ref{fig:endres_vs_si} shows the dependence of resolution based on the video's SI for different personality types. The dependence is the same regardless of the personality type (all have the same slope), however, the intercept is different. This is in line with the results of the linear regression model (Table~\ref{tab:lrm_pers2}) and also of the boxplot distribution of end resolutions for each dominant personality type (Figure~\ref{fig:boxplot_personality}): agreeableness requires the highest overall resolutions, while other traits exhibit similar behavior, with openness requiring the lowest overall resolution. Individuals who score high on agreeableness tend to be compliant and cooperative, and to conform with rules not to upset others~\cite{roccas2002big,graziano1997agreeableness}. In our study, viewers with agreeableness as dominant personality trait might have focused on the task of changing resolution as their goal in this experiment, and thus have been more keen on changing the resolution in order to satisfy the requirements of the study.

\begin{figure}
  \includegraphics[width=\columnwidth]{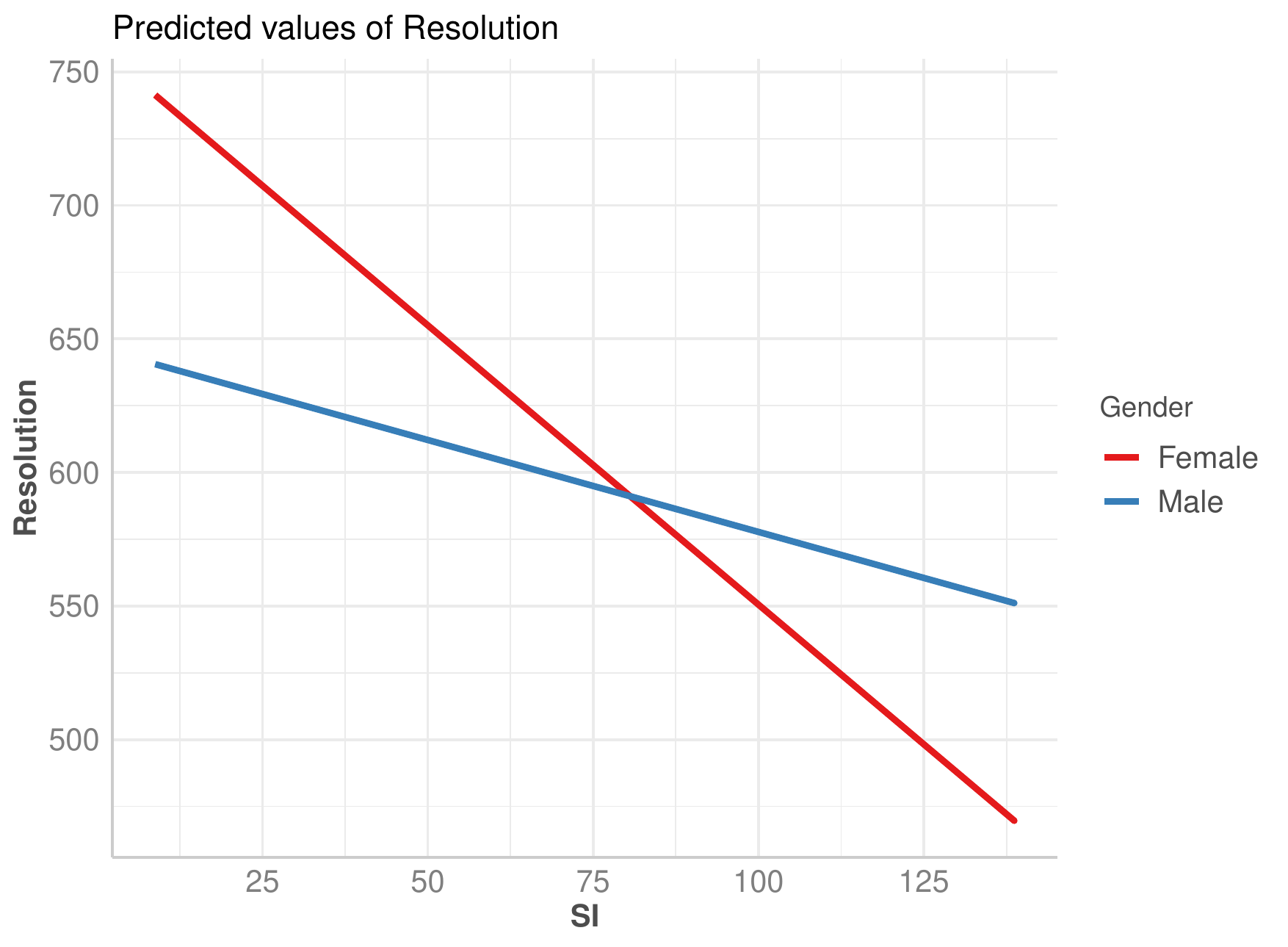}
  \caption{End resolution vs. SI for female vs. male viewers. While the slope is negative in both cases, the decrease for female viewers is more steep as SI increases compared to male viewers.}
  \label{fig:gender}
\end{figure}

\begin{figure}
  \includegraphics[width=\columnwidth]{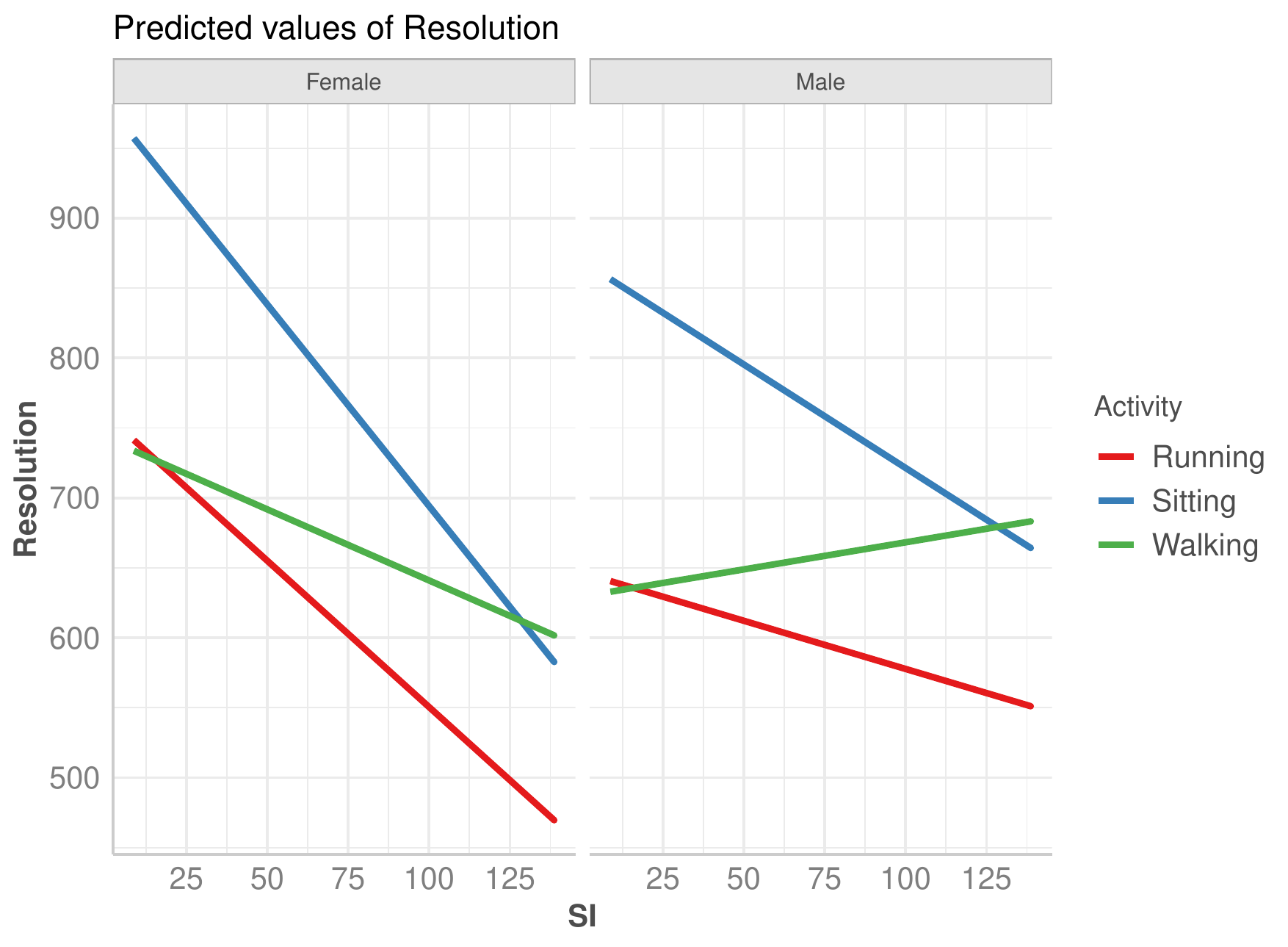}
  \caption{End resolution vs. SI for male vs. female viewers in each activity state. This chart confirms that for videos with higher SI, female viewers require lower resolutions than male viewers. Also, walking state stands out as male viewers require higher resolutions for higher SI in this state, a reverse effect than the one encountered for female viewers. }
  \label{fig:gender_activity}
\end{figure}

The impact of gender on end resolution is illustrated in Figure~\ref{fig:gender}. This plot shows that as the SI increases all viewers require lower resolutions. However, the slope is different for male vs. female, with male viewers requiring higher resolutions than female viewers for high SI videos. This trend is also confirmed in Figure~\ref{fig:gender_activity}, when visualizing SI vs. resolution for different activities for each gender. The slopes for female viewers are more steep than for male viewers, and the intercepts for male viewers are higher than for female viewers. For all activities, male viewers require lower resolutions than female viewers for low SI videos. However this trend decreases as the SI of the video increases, and for high SI videos it reverses. In addition, when walking male viewers require higher resolutions as the SI increases, while female viewers require lower. 

This unusual observation could be explained by a difference in interest of male vs. female viewers for the content of the highest SI videos in the selection (as illustrated in Figure~\ref{fig:study2_thumbnails}). Related literature has highlighted that gender, among other factors, plays an important role in the interest in a particular video content~\cite{INFORTUNA2021110877}. In our video selection, the highest scoring SI videos comprise sports, online video tutorial, and animated sketches. 


\subsection{Predictive Context- and Personality-Aware Mobile Video Resolution Model}
\label{sec:results:predictive}

The statistical and hierarchical analysis performed showed that the Activity, SI and Gender, together with their interactions, and also the dominant personality trait, all impact the viewer’s quality requirements when watching videos on a mobile device. Based on these results, we want to be able to predict from the mobile sensed data how to best adapt the resolution. As such, we now move a step further and construct machine learning models that take these parameters at the input and predict the most suitable viewing resolution.

First, we train two regressors: a Random Forest regressor and a mean regressor to serve as a baseline (a regressor which always predicts the mean of training target values). We employ the Leave-One-Out Cross-Validation (LOOCV) procedure, a specific type of k-fold cross validation, where the number of folds, k, is equal in our case to the number of viewers in the dataset. As such, each time we train the model on the data from 22 viewers and test it on the “left out” viewer. For each “fold” we compute the following accuracy metrics: prediction accuracy (using the mean average percentage error subtracted from 100\%), mean absolute error (MAE) and root mean squared error (RMSE). Finally, to assess the performance of the entire model, we take the mean and standard deviation of these accuracy metrics. The results we obtained are illustrated in Table~\ref{tab:regressor1}.

\begin{table}
\centering
\caption{Prediction performance comparison of the Random Forest regressor vs. a mean regressor.}
\label{tab:regressor1}
\begin{tabular}{lllllll}
   \toprule
Method         & Av. Acc & Std. & Av. MAE & Std. & Av. RMSE & Std. \\
\midrule
Random Forest  & 73.7\%  & 11.6 & 159.7   & 67.3 & 201.7    & 74.8 \\
Mean Regressor & 67.6\%  & 5.8  & 192.4   & 42.2 & 225.8    & 49.6\\
\bottomrule

\end{tabular}
\end{table}

These numbers show that on average, the Random Forest regressor achievies an accuracy of $73.7\%$ in predicting the appropriate viewing resolution, higher than the $67.6\%$ accuracy scored by the mean regressor. The MAE and RMSE values are also better for the Random Forest regressor, however, for all 3 performance metrics the standard deviation values are higher compared to the ones of the mean regressor. This indicates that there are significant differences in the accuracy of the predictions varying from viewer to viewer. 

Motivated by these differences, we next proceed to build dedicated predictive models for each of the dominant personality traits. We exclude Conscientiousness since the dataset contains only one viewer with this dominant personality trait. Similiarly, we use the LOOCV procedure, and for each dominant personality we build a Random Forest regressor and a mean regressor. We compute the same accuracy metrics and the results are shown in Table~\ref{tab:regressor2}.

\begin{table}
\centering
\caption{Prediction performance comparison of the Random Forest regressors vs. mean regressors for each dominant personality trait.}
\label{tab:regressor2}
\begin{tabular}{clllllll}
   \toprule
Personality           & Method         & Av. Acc & Std.  & Av. MAE & Std. & Av. RMSE & Std. \\
\midrule
\multirow{2}{*}{Agreeableness} & Random Forest  & 54.0\%  & 23.2  & 269.1   & 71.7 & 323.4    & 83.8 \\
                               & Mean Regressor & 63.7\%  & 11.2  & 219.8   & 48.2 & 267.8    & 45.0 \\
\multirow{2}{*}{Extraversion}  & Random Forest  & 78.2\%  & 4.0   & 156.1   & 54.3 & 207.5    & 74.3 \\
                               & Mean Regressor & 68.1\%  & 4.7   & 192.2   & 33.0 & 226.6    & 42.5 \\
\multirow{2}{*}{Neuroticism}   & Random Forest  & 77.5\%  & 9.6 & 136.0   & 49.9 & 177.3    & 54.6 \\
                               & Mean Regressor & 67.9\%  & 6.2  & 191.0   & 31.7 & 229.1    & 35.6 \\
\multirow{2}{*}{Openness}      & Random Forest  & 81.1\%  & 4.8   & 114.5   & 41.2 & 161.8    & 47.5 \\
                               & Mean Regressor & 69.8\%  & 2.3   & 170.2   & 51.8 & 192.1    & 78.0 \\
\bottomrule
\end{tabular}
\end{table}

The results show that for 3 out of 4 personality types the personality-specific Random Forest regressors achieve higher prediction accuracy than the general Random Forest model, with Agreeableness being an exception (likely due to limited dataset, which might also generally explain the limited performance of all the Random Forest regressors). However, the issue of high values for the standard deviation for all accuracy metrics remains, indicating that using solely these parameters the model fails to fully adapt to individual behavior. This confirms the findings of the statistical and hierarchical analysis presented in sections~\ref{sec:results:properties} and~\ref{sec:results:hierarchical}, which indicated there are additional viewer-related factors that impact the quality requirements and which require future investigation.
\section{Related Work}
\label{sec:related_work}

\subsection{Energy-efficient mobile multimedia}

The limited battery charge became the key pressing issue preventing further growth of mobile computing~\cite{1401839} and exacerbating the need for utilizing the available resources as efficiently as possible. 
Among the services consuming the largest amount of energy in a mobile device, multimedia apps \cite{shin2019reducing,lbu4703} stand out, together with network traffic \cite{yan2019modeling} and machine learning \cite{mcintosh2019can}. Yet, the high popularity of mobile multimedia makes addressing the energy consumption of such apps a pressing issues. A recent Atos study \cite{atos2019} reveals that mobile multimedia apps are the second most intensively used applications (based on average time spent by the user) and consequently also rank second in impact on the average daily energy consumption of a mobile device. 

Solutions for reducing the energy consumption of mobile video apps include the work by Shin et al.~\cite{shin2019reducing}, where the authors present an approach for reducing the energy consumption of random network coding based media streaming applications on smartphones by manipulating the frequency controllers in the smartphone's operating system. Another solution proposed by Hu and Cao~\cite{7980188} introduces an energy-aware CPU frequency scaling algorithm for mobile video streaming, which selects the CPU frequency that can achieve a balance between saving the data transmission energy and CPU energy. Ahmad et al.~\cite{ahmad2018battery} developed a battery-aware rate adaptation for extending video streaming playback time which adapts to the appropriate bit rate to prolong the battery lifetime. An energy efficient video decoding for the Android operating system is proposed by Liang et al. \cite{liang2013energy}, based on dynamic voltage and frequency scaling. Hamzaoui et al. in their work \cite{hamzaoui2020measurement} propose a measurement-based methodology for modeling the energy consumption of mobile devices and use video decoding tasks (both on-device and remote streaming) for the experimental power measurements. 

Most of the above-mentioned energy-saving solutions focus on optimizations at the hardware and network layer for video streaming; by comparison, our approach is hardware-agnostic and adapts the video resolution according to the user's context, which influences his quality requirements. In addition, this context- and content-aware adaptation strategy has the advantage of being applicable for both network video streaming and on-device playback.  

\subsection{Mobile video quality perception}

Perception of multimedia quality is impacted by a synergy between system, context and human factors \cite{scott2016personality}. The continuous technological advances in multimedia services have enabled them to be increasingly optimized in a personalized way, by taking into account the human factors when estimating the Quality-of-Experience (QoE) in order to optimize the video delivery to the user \cite{zhu2015towards,zhu2018measuring}. Hence, numerous research efforts have been carried out to analyze the influence of system, contextual and human factors on the perception of multimedia quality \cite{zhu2015understanding,scott2015modelling,satgunam2013factors}.

Dynamic viewing environment makes mobile video strikingly different from the conventional TV or Desktop PC viewing experience. Contextual factors, such as whether a viewer is indoor or outdoor, walking, running or riding a bus, and others, may change even during a single viewing session~\cite{6777577}. Research in this field identified several factors that influence mobile video quality perception, such as the display size, viewing distance from the display, environmental luminance, and physical activity of the user and showed that environment-aware video rate adaptation can enhance mobile video experience while reducing the bitrate requirement by an average of 30\% \cite{6777577}. Another study shows that in the mobile environment, sensory experience is a significant factor for enjoyment and engagement with the video as outside interruptions decrease the user's video quality experience on a mobile device~\cite{SEETO20121484}. This might be the reason for heavy tailed distributions of selected resolutions when users are walking or running, observed in our dataset. It is possible that, while generally too distracted to pay attention to fine video details, at certain occasions, users select a higher resolution to counter the effect of environmental disruptions.

The correlation between video content and user perceptual satisfaction is underlined by the existing research focused on this phenomena. Trestian et al. demonstrate a low spatial information video watched in low quality is likely to be found more acceptable/satisfying by the user than watching a high spatial and temporal complexity video the same quality~\cite{trestian}. The research findings also support the theory that one can expect significant differences in the user satisfaction at the same quality level depending on the particularities of the video. More specifically, the authors observed 20\% average user satisfaction level difference between two videos watched in the lowest quality setting. We can see this in our study as well: from a subset of videos watched by users in ``still'' and ``running'' states, the video with a very high spatial complexity stands out as requiring a substantially higher resolution from the users when running, compared to the other videos in the subset which had lower SI scores (Figure~\ref{fig:still_vs_running}). This indicates that the the video's spatial information feature influences the user's quality expectations in physically active states, such as running.

Song et al. identify a stronger relationship between acceptability and content type at a relatively low bitrate range of 200 -- 400kbps~\cite{Song2011}. The paper also concludes that the acceptability rate is influenced by the video content type: at higher resolutions, such as 480x320 pixels and 640x480 pixels, acceptability higher than 60\% can be achieved, if the bitrate is greater than 300 Kbps for news, 400 Kbps for animation, movie, and music, and 800 Kbps for sports videos. The video content directly impacts the video's spatial and temporal information scores, e.g. animations usually have lower SI/TI, while sport videos have much higher scores. Our study confirms this: the videos with the highest SI and TI are either sport videos (basketball match -- video 2, car dashboard camera recording -- video 11 or body camera recording of mountain bike trail -- video 5).

In \cite{scott2016personality} and \cite{scott2015modelling} the authors studied the interplay between system, context, and human factors on the perceived video quality and enjoyment. Both studies showed that human factors play an important role in the way perception of quality and enjoyment are rated. In addition, the nature of the content alone is more likely to influence how it is perceived than the system settings at which it is delivered.The analysis of perceptual quality, in particular, indicated that a greater proportion of the variance can be predicted by human factors (24.3\%) than by system factors (13.7\%); however, all the system factors and most of their interactions have larger effect sizes than any individual human factor. This implies that perceived quality and enjoyment are determined by humans as much as they are determined by multimedia systems. The studies have examined closely two measurable dimensions of the human factors, namely the personality and cultural traits. The statistical results indicated that these two factors represent a small portion of the variance which can be attributed to human factors. Collectively, both sets of variables represent 9.3\% of the variance, indicating that there are additional, more crucial human-related dimensions that impact the perceived quality and enjoyment.

The question regarding how exactly the user's personality (and which of its dimensions) impacts the quality and enjoyment perception of multimedia content is debated among researchers. In an earlier study on this topic, Gulliver and Ghinea~\cite{gulliver2010cognitive} distinguish three dimensions of the overall user satisfaction with a video (the overall Quality of Perception -- QoP): level of enjoyment (QoP-LoE), level of information the users believe they assimilated (QoP-LoA), and the level of confidence the user has with regard to the information assimilated (QoP-LoC). In their study, they used the Myers–Briggs Type Indicator (MBTI) to describe the participants' personalities. They concluded that among the three dimensions of a user's satisfaction with a video, no significant results were found between personality dimensions and QoP-LoE. In the same time, personality dimensions significantly affected user self-perceived QoP-LoA and QoP-LoC. Their conclusion seems to be confirmed by another study by Satgunam et al.~\cite{satgunam2013factors}, where the authors investigated factors affecting enhanced video quality preferences and find that while human factors play an important role overall, personality did not seem to relate with the video enhancement preferences. They concluded this after administering a personality questionnaire related to the tolerance of blur.

In the work by Zhu et al.~\cite{zhu2018measuring}, the authors present their study on the individual factors influencing video QoE (Quality of Experience), conducted using an open-source Facebook application developed for this purpose, named YouQ. Their results are presented and compared with other two studies that investigated systematically the influence of user factors on individual Quality of Experience~\cite{zhu2016qoe,scott2015modelling}. The three-way comparison shows that all three studies confirm the importance of user factors since a large proportion of variance can be explained by considering users as a “random effect”, especially on the results of YouQ. With regard to the correlation between the personality (all three studies used the Big five personality traits model) and the user enjoyment and quality perception, the results were mixed: regarding the influence on perceived quality, YouQ found no significant relationship, i\_QoE~\cite{zhu2016qoe} found that a user who has a more agreeable personality tends to rate the perceived quality significantly higher, while CP-QAE-I~\cite{scott2015modelling} that a user who is conscientious rates perceived quality of a video significantly more.

\section{Conclusions}
\label{sec:conclusions}

In this work we assessed the feasibility of dynamic energy efficient context-aware mobile video playback adaptation, employing an approach fostered by the philosophy of approximate mobile computing. After showing that playing videos on mobile devices at higher quality (resolution) increases the energy consumption, we hypothesised that the actual viewer quality expectations are not constant in the mobile environment, but instead vary with the ``context''. To explore the potential dimensions of the context in mobile video playback, we started by conducting an initial user experience study which revealed that the resolution found acceptable by viewers was influenced by the physical activity state of the viewer, and also the video content, more specifically its spatial and temporal characteristics. In addition, this study showed that there are other viewer-related factors (e.g. personality, cultural background) that may impact a user's perception of a mobile video playback.

As such we conducted a second user experience study, involving 23 participants and which was focused on gathering additional information about the user's personality traits. We examine the data of this study by both simple statistical analysis and mixed effects modelling to take into account not just the fixed effects of the parameters but also the nested nature of our data (i.e. grouped by personality type). Such a detailed analysis demonstrates that a viewer's mobile multimedia quality expectations indeed exhibit significant context-dependent variations. These variations, however, remain rather nuanced, lightly steered by different contextual, content, and viewer-related factors. More specifically, we find that:
\begin{itemize}
    \item A viewer's physical activity, in general, negatively impacts the desire for a higher video resolution. As a consequence, a simple resolution adaptation driven by automatic activity detection represents a low-hanging fruit for energy-efficient video playback.
    \item Spatial and temporal properties of a video impact the desired resolution, yet often only when a viewer is on the move. The impact, however, remains subtle and difficult to disentangle from other factors. In our first study, for instance, we find that the viewers require a higher resolution for high-SI videos when running.
    \item A viewer's dominant personality may impact the required playback resolution. Observing that the highest resolution is selected by agreeable viewers, we hypothesise that this is due to these viewers' desire to comply with the presumed goals of the study and indulge the researchers~\cite{yoursIsBetter}.
    \item A viewer's interest in the topic of a video may drive the desire for a higher resolution in certain contexts. While in this work we do not explicitly measure such desire (e.g. through interviews with the participants), we observe that a viewer's gender, as a weak proxy for the interests, drives the desired resolution when videos of different spatial information are watched. 
\end{itemize}



After uncovering these factors, we moved to assess the feasibility of machine learning models that predict the acceptable final resolution. We trained a general Random Forest regressor using the Leave-One-Out Cross-Validation strategy and evaluated it using several accuracy metrics. The model achieved an average accuracy of 73.7\% (c.f. 67.6\% baseline), but experienced high variations in the prediction accuracy among viewers. To take into account the differences in viewer preferences influenced by their personality traits, we then elaborated separate personality-specific regressors, which in general achieved better prediction accuracies than the generic prediction model, but still revealed the limitations due to the small dataset size. This indicates that the activity information, SI, TI, and personality traits may not be sufficient for training a generally applicable machine learning model for mobile video resolution adaptation. In future work we plan to examine incremental and transfer learning in order to tune the model to individual users.

Our research represents the initial step demonstrating the link between the mobile multimedia quality expectations and the context of use. Importantly, we show that even with readily available information (i.e. activity, SI/TI) and tools (video resolution dial) we can already enable energy savings, thus address the critical issue of constrained battery capacity in mobile devices. Assessing the amount of energy savings achievable via mobile video resolution adaptation is outside of the scope of our work, as it requires the knowledge of the actual distribution of parameters (SI and TI) of mobile videos viewed by a user and the context (e.g. activity) in which videos are watched. Finally, our work indicates that aside from the viewer's physical activity, the content of the video and the viewer's personality, there are also other dimensions that impact the quality requirements and which must be further explored in order to enable accurate prediction of the appropriate quality settings for video playback. To facilitate further research along this front we make our study data available at \url{https://gitlab.fri.uni-lj.si/lrk/approximate_video_study/}.

%
%

\bibliographystyle{spmpsci}      
\bibliography{references}   

\end{document}